\PassOptionsToPackage{table}{xcolor}
\documentclass{article}

\usepackage[preprint]{neurips_2026}

\usepackage[utf8]{inputenc} 
\usepackage[T1]{fontenc}    
\usepackage{url}            
\usepackage{booktabs}       
\usepackage{amsfonts}       
\usepackage{nicefrac}       
\usepackage{microtype}      
\usepackage{xcolor}         
\usepackage{multirow}
\usepackage{pifont}
\usepackage{graphicx}
\usepackage{subcaption}
\usepackage{amsmath}

\usepackage{tabularx}
\usepackage{enumitem}
\usepackage{amssymb}
\usepackage{comment}
\usepackage{hyperref}       

\usepackage{wrapfig}
\usepackage{tikz}
\usetikzlibrary{fit,calc,positioning}
\usepackage{siunitx}
\setcitestyle{numbers}
\usepackage{soul}

\title{X-Palm: Paired Multispectral-to-Smartphone Dataset for Cross-Domain Palmprint Authentication}

\author{%
  Jamal Seyedmohammadi \\
  Singapore Institute of Technology \\
  \And
  Pai Chet Ng \\
  Singapore Institute of Technology \\
  \And
  Angelo Genovese \\
  Universit\`a degli Studi di Milano \\
  \AND
  Zhixiang Chi \\
  University of Toronto \\
  \And
  Jeannie Lee \\
  Singapore Institute of Technology \\
  \And
  Konstantinos N. Plataniotis \\
  University of Toronto \\
}

\begin{document}

\maketitle

\begin{abstract}
Palmprint modality offers a privacy-preserving biometric solution, yet its deployment is hindered by the domain gap between controlled enrollment and unconstrained authentication. Existing datasets are largely restricted to controlled setups and fail to capture the compound variability of real-world environments.
In this paper, we introduce \textbf{X-Palm}, a cross-domain dataset comprising 6,006 palm images from 103 individuals (206 hands). 
To the best of our knowledge, X-Palm is the first palmprint dataset providing novel paired-identity acquisition specifically designed to bridge the gap between reliably controlled multispectral enrollment and unconstrained mobile authentication while encompassing a broad spectrum of in-the-wild variability.
Unlike existing datasets that focus on single to a few variations, X-Palm addresses the massive modality and environmental shifts encountered in practical deployments by capturing paired data for identities across two distinct domains:
(1) a controlled Multispectral Palmprint setting using our custom-developed scanner, and
(2) an unconstrained smartphone palmprint setting that is participant-driven, incorporating simultaneous variations in hardware, hand pose, illumination, background, camera-to-hand distance, perspective, and palm surface conditions (e.g., moisture and occlusions). 
Our extensive benchmarks of 12 SOTA models reveal that while existing methods achieve high performance on controlled data, they experience severe performance collapse on X-Palm. 
Conversely, models trained on X-Palm demonstrate consistent robustness across domains, positioning X-Palm as a valuable resource for training a model towards real-world, cross-domain generalization.
Data access instructions and the related benchmarking codes are publicly available at: \url{https://github.com/X-Palm/X-Palm-2026}

\end{abstract}

\section{Introduction}
\begin{wrapfigure}{r}{0.5\textwidth} 
\vspace{-0.2cm}
  \centering
  \includegraphics[width=0.54\textwidth]{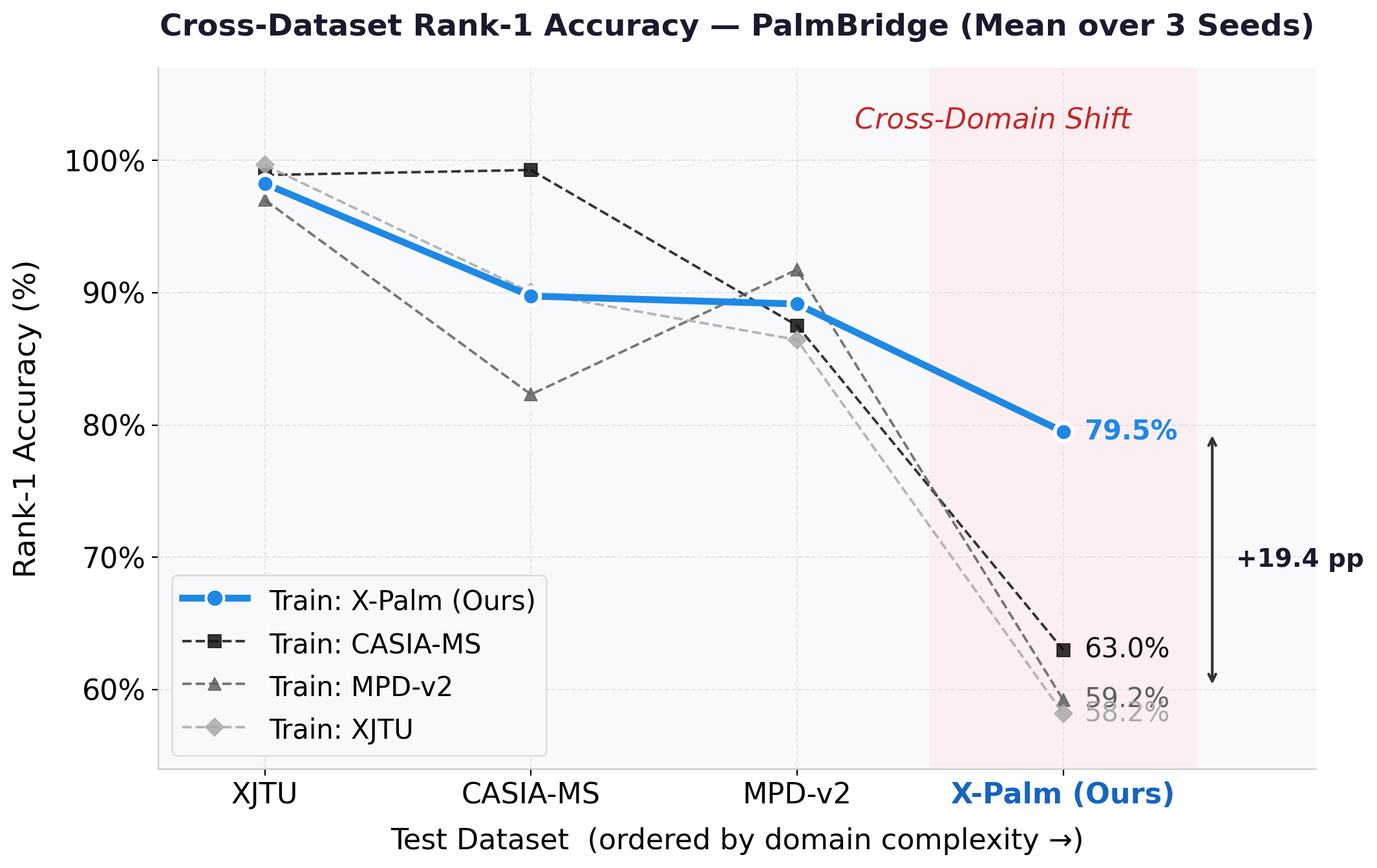}
  \caption{The Generalization Gap: Models trained on existing datasets generalize poorly under cross-domain shift, while training on X-Palm improves robustness, motivating paired multispectral-to-smartphone data.}
  \label{fig:motivation_gap}
  \vspace{-0.3cm}
\end{wrapfigure}

Palmprint-based biometrics has emerged as a reliable and privacy-preserving authentication modality due to the rich discriminative information embedded in the human palm, including principal lines, wrinkles, creases, and fine-grained textural patterns~\cite{11353238, yang2023multi}. With the increasing use of biometric authentication in human identification~\cite{5783341}, mobile banking~\cite{yu2026show}, intellectual property protection~\cite{9722708}, and contactless access~\cite{10555397}, palmprint recognition offers a practical solution that combines user convenience with strong identity discrimination. However, real-world deployment remains challenging because palmprint systems must generalize across domains: from controlled enrollment to unconstrained authentication. As illustrated in Fig.~\ref{fig:motivation_gap}, this cross-domain gap causes models trained on clean, controlled data to suffer severe performance degradation when faced with real-world palm images captured under different devices, lighting conditions, viewpoints, and user behaviors. 

\textbf{Controlled Enrollment Challenge.} 
In many high-security authentication scenarios, the initial enrollment stage is expected to produce a reliable biometric template~\cite{9815514}. Therefore, enrollment is often performed under controlled acquisition conditions using dedicated hardware, fixed illumination, stable imaging distance, and guided hand placement. Existing multispectral palmprint datasets, such as CASIA-MS~\cite{casiams}, are collected under controlled environment and are valuable for studying recognition under such clean conditions. However, these datasets mainly represent the enrollment domain and do not reflect the conditions under which users will later authenticate in daily use. Models trained only on controlled enrollment data may therefore overfit to clean acquisition settings and fail when the authentication image is captured by a different device under less constrained conditions.

\textbf{Unconstrained Authentication Challenge.} 
In practical mobile authentication, users are more likely to verify their identity using personal smartphones in everyday environments~\cite{10752516}. This introduces compound variability, where multiple sources of domain shift occur simultaneously. Palm images may vary due to smartphone sensor differences, illumination changes, hand pose, camera-to-hand distance, perspective distortion, background clutter, motion blur, occlusion, and palm surface conditions such as moisture or handwritten marks. Existing smartphone palmprint datasets, such as MPD-v2~\cite{mpd} and XJTU-UP~\cite{xjtu}, introduce device and lighting variations, while NTU-PI~\cite{ntu} provides more unconstrained Internet-collected images. Nevertheless, these datasets either lack paired controlled enrollment data for the same identities or cover only limited variations per subject. 

\textbf{X-Palm.} 
As summarized in Table~\ref{tab:dataset_comparison}, existing datasets capture only partial aspects of the real-world deployment challenge. 
To address these limitations, we introduce \textbf{X-Palm}, a paired multispectral-to-smartphone palmprint dataset designed specifically for cross-domain palmprint authentication. 
X-Palm is not merely another palmprint dataset with additional variations; its key novelty lies in its paired data design. For the same identities, X-Palm provides palm images from two distinct domains: a controlled multispectral enrollment domain captured using our custom-developed scanner, and an unconstrained smartphone authentication domain captured through participant-driven mobile collection. This paired structure enables direct study of the domain gap between high-trust enrollment and real-world authentication. In addition, the smartphone collection protocol deliberately captures compound variability, including differences in device, lighting, hand pose, distance, perspective, background, occlusion, and palm surface condition.

\begin{table*}[t!]
  \centering
  \caption{Comparison of palmprint datasets based on variations and features. 
  Pos: hand pose; Lig: lighting; Dis: distance; Per: perspective; Occ: occlusion; 
  Bac: background; Dev: device; All: all variations for each identity; 
  CSA: cross-setting authentication.}
  \label{tab:dataset_comparison}
  \resizebox{\textwidth}{!}{
  \begin{tabular}{l ccccccccc cccc c}
    \toprule
    \multirow{2}{*}{\textbf{Datasets}} 
    & \multicolumn{9}{c}{\textbf{Variations}} 
    & \multicolumn{5}{c}{\textbf{Features}} \\
    \cmidrule(lr){2-10} \cmidrule(lr){11-15}
    & \textbf{Pos.} 
    & \textbf{Lig.} 
    & \textbf{Dis.} 
    & \textbf{Per.} 
    & \textbf{Occ.} 
    & \textbf{Bac.} 
    & \textbf{Dev.} 
    & \textbf{All} 
    & \textbf{CSA} 
    & \textbf{\#hands} 
    & \textbf{\#img/hand} 
    & \textbf{\#img} 
    & \textbf{\#Device} 
    & \textbf{Age Range} \\
    \midrule
    CASIA-MS~\cite{casiams}  
    & \ding{55} & \checkmark & \ding{55} & \ding{55} & \ding{55} 
    & \ding{55} & \ding{55} & \ding{55} & \ding{55} 
    & 200 & 36 & 7,200 & 1 & 20--63 \\

    XJTU-UP~\cite{xjtu}   
    & \ding{55} & \checkmark & \ding{55} & \ding{55} & \ding{55} 
    & NA & \checkmark & \ding{55} & \ding{55} 
    & 200 & 100 & 20,000 & 5 & 19--35 \\

    MPD-v2~\cite{mpd}   
    & \ding{55} & \checkmark & \ding{55} & \ding{55} & \ding{55} 
    & \checkmark & \checkmark & \ding{55} & \ding{55} 
    & 400 & 40 & 16,000 & 2 & 20--50 \\

    NTU-PI~\cite{ntu}    
    & \checkmark & \checkmark & \checkmark & \checkmark & \checkmark 
    & \checkmark & NA & \ding{55} & \ding{55} 
    & 2,035 & 4 & 7,781 & NA & NA \\

    \rowcolor{gray!15}
    X-Palm (ours) 
    & \checkmark & \checkmark & \checkmark & \checkmark & \checkmark 
    & \checkmark & \checkmark & \checkmark & \checkmark 
    & 206 & 18 or 33 & 6,006 & 80+ & 18--76\textsuperscript{\dag} \\
    \bottomrule
  \end{tabular}
  }
  \begin{flushright}
  \tiny
  \textsuperscript{\dag}X-Palm includes 103 participants with 40.78\% female subjects and 11 self-reported ethnic groups.
  \end{flushright}
\end{table*}

\textbf{Contributions.} 
Our contributions are two-fold. 
First, our \textbf{X-Palm}, to the best of our knowledge, is the first palmprint dataset that pairs controlled multispectral acquisition with unconstrained smartphone acquisition from the same identities. 
X-Palm contains 6,006 palm images from 103 participants and 206 hands, with the large-scale compound variability in illumination, hand pose, distance, perspective, background, occlusion, and palm surface condition, together with broad demographic coverage. 
Second, we provide a comprehensive benchmark for cross-domain palmprint authentication. We evaluate 12 representative models under cross-dataset, closed-set cross-domain, and open-set cross-domain protocols, revealing that existing models trained on prior datasets degrade substantially on X-Palm, whereas models trained on X-Palm achieve stronger average cross-dataset generalization.

\begin{figure}[t]
    \centering
    \includegraphics[width=\textwidth]{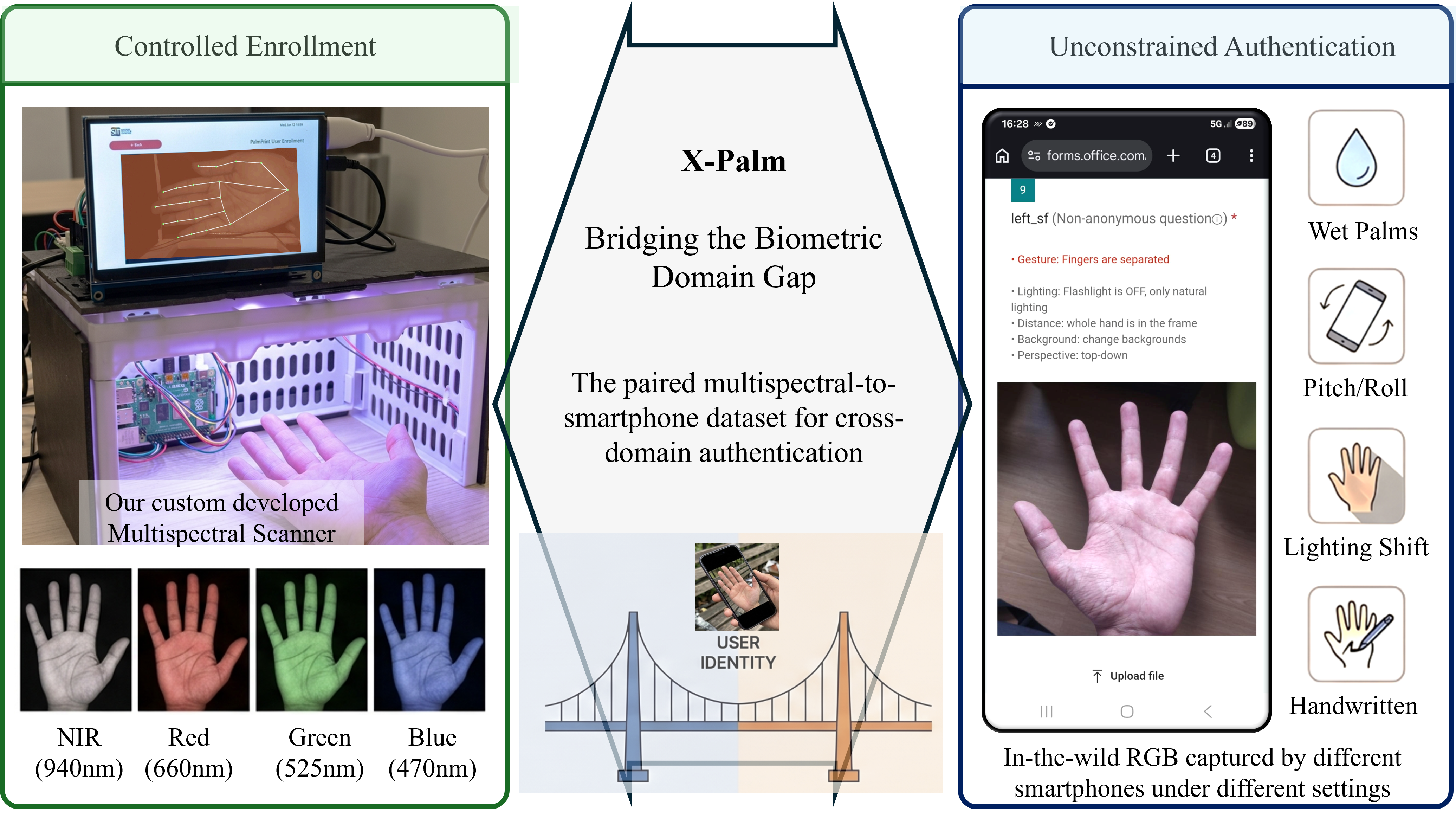}
    \caption{Overview of X-Palm Data Collection: The collection protocol is carefully designed to bridge the gap between controlled multispectral enrollment and unconstrained smartphone authentication, while capturing compound in-the-wild variability. }
    \label{fig:setup}
\end{figure}
\section{X-Palm Data Curation and Preparation}
This section outlines the methodology we employed to collect palm images and provides a detailed description of our carefully curated dataset. 

\subsection{Paired Multispectral-to-Smartphone Data Collection Framework}
Our novel paired multispectral-to-smartphone data collection framework, as illustrated in Fig.\ref{fig:setup}, is designed to capture the same identity data across two distinct domains: a reliably controlled multispectral scanner and unconstrained smartphone.

\textbf{Controlled Multispectral Environment. }
In this setting we use our custom-developed, multispectral scanner. 
As shown in Fig.~\ref{fig:setup}(left),  the device integrates a programmable illumination module, low-cost RGB and IR cameras, and a Raspberry Pi control hub. 
At the hardware level, the DMX-controlled LED module emits multiple spectral channels, including red, green, blue, white, yellow, and infrared, to acquire high-resolution palmprint images at $4608\times2592$. 
The Raspberry Pi serves as the orchestration layer, synchronizing lighting sequences and camera exposures while hosting a dedicated user interface that allows for real-time monitoring and quality verification of captured templates~\footnote{Check out our custom-developed multispectral scanner here: \url{https://youtu.be/7QgTzKBEeKE}}. 

\textbf{Unconstrained Mobile Authentication. }
In contrast to the controlled scanner, the smartphone setting utilizes an unconstrained, participant-driven protocol leveraging the subjects' personal smartphones.
As shown in Fig.~\ref{fig:setup}(right), data collection is facilitated through a custom-configured Microsoft Form Application, serving as both an instructional guide and a secure submission portal.
This setup is explicitly designed to capture simultaneous compound variability, including variations across different smartphone brands and models, hand poses, illumination, and palm surface conditions (e.g., moisture and occlusions), to simulate the unpredictability of real-world mobile deployments.

\begin{figure*}[t]
    \centering
    \captionsetup[subfigure]{font=scriptsize}
    \begin{tikzpicture}[inner sep=5pt]
        
        \node (controlled) {
            \begin{minipage}{\textwidth}
                \centering
                \begin{subfigure}{0.15\textwidth}
                    \includegraphics[width=\linewidth]{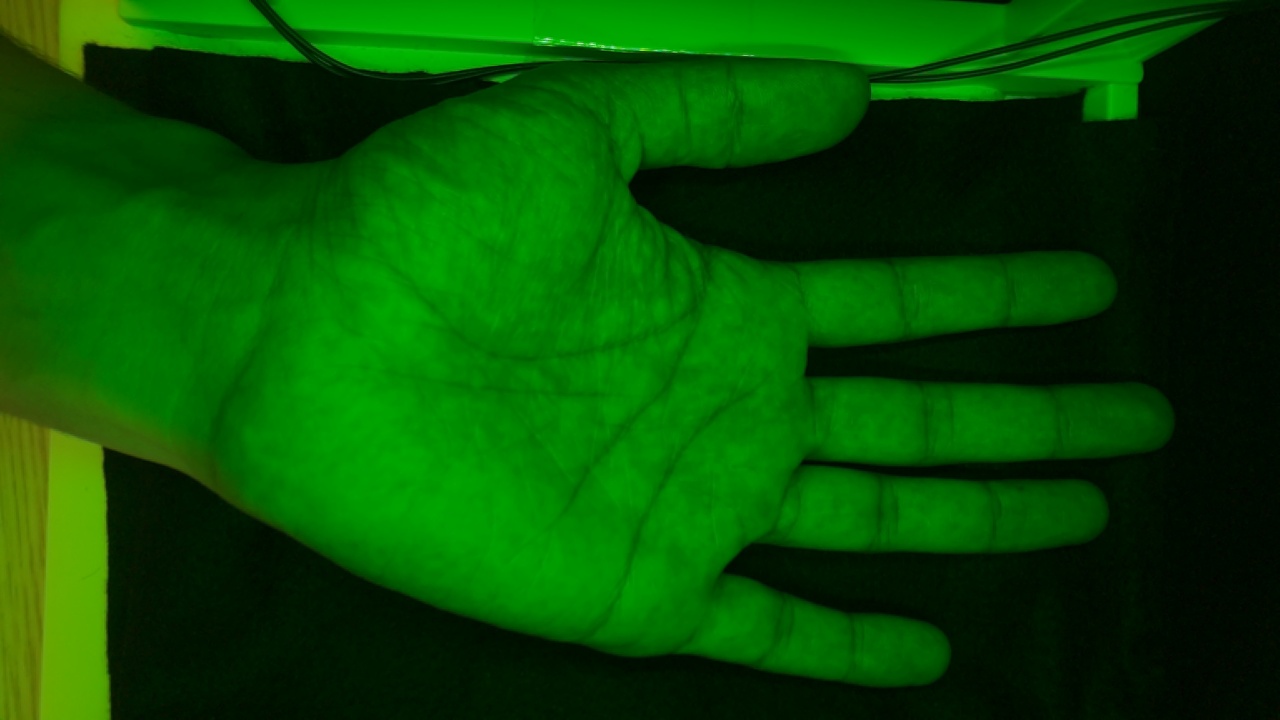}
                    \caption{Green}
                \end{subfigure}\hfill
                \begin{subfigure}{0.15\textwidth}
                    \includegraphics[width=\linewidth]{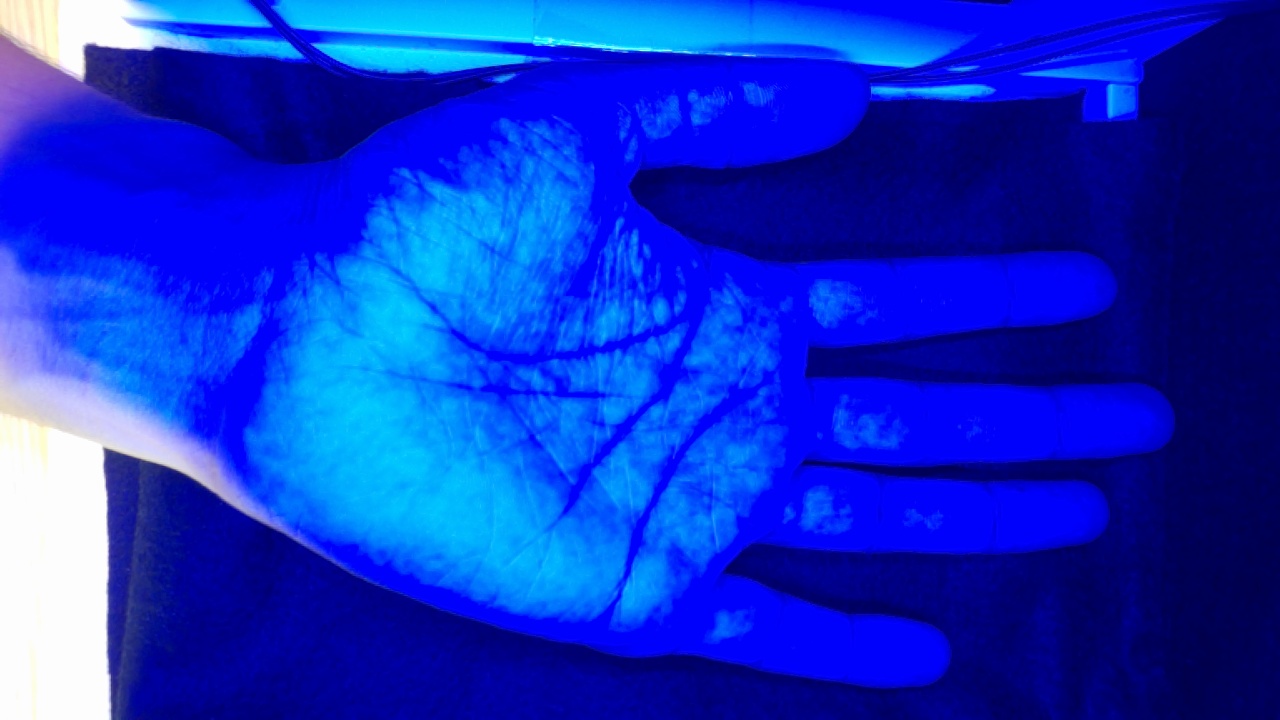}
                    \caption{Blue}
                \end{subfigure}\hfill
                \begin{subfigure}{0.15\textwidth}
                    \includegraphics[width=\linewidth]{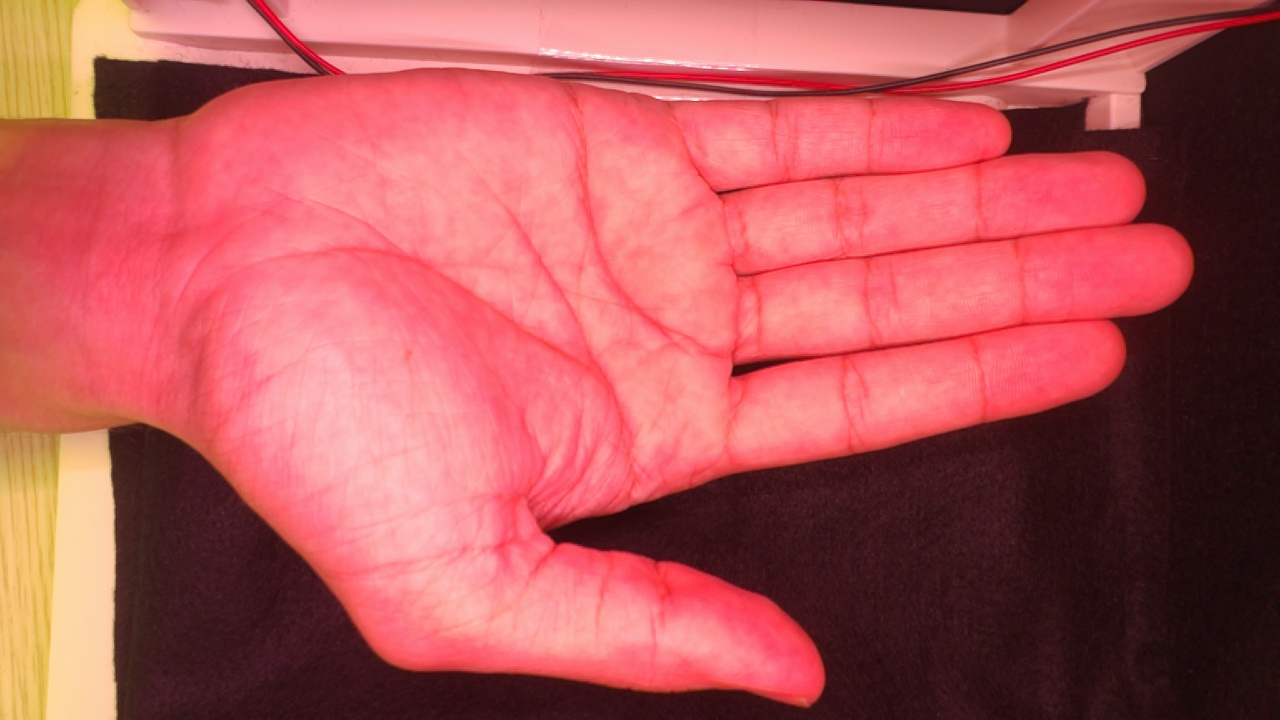}
                    \caption{Red}
                \end{subfigure}\hfill
                \begin{subfigure}{0.15\textwidth}
                    \includegraphics[width=\linewidth]{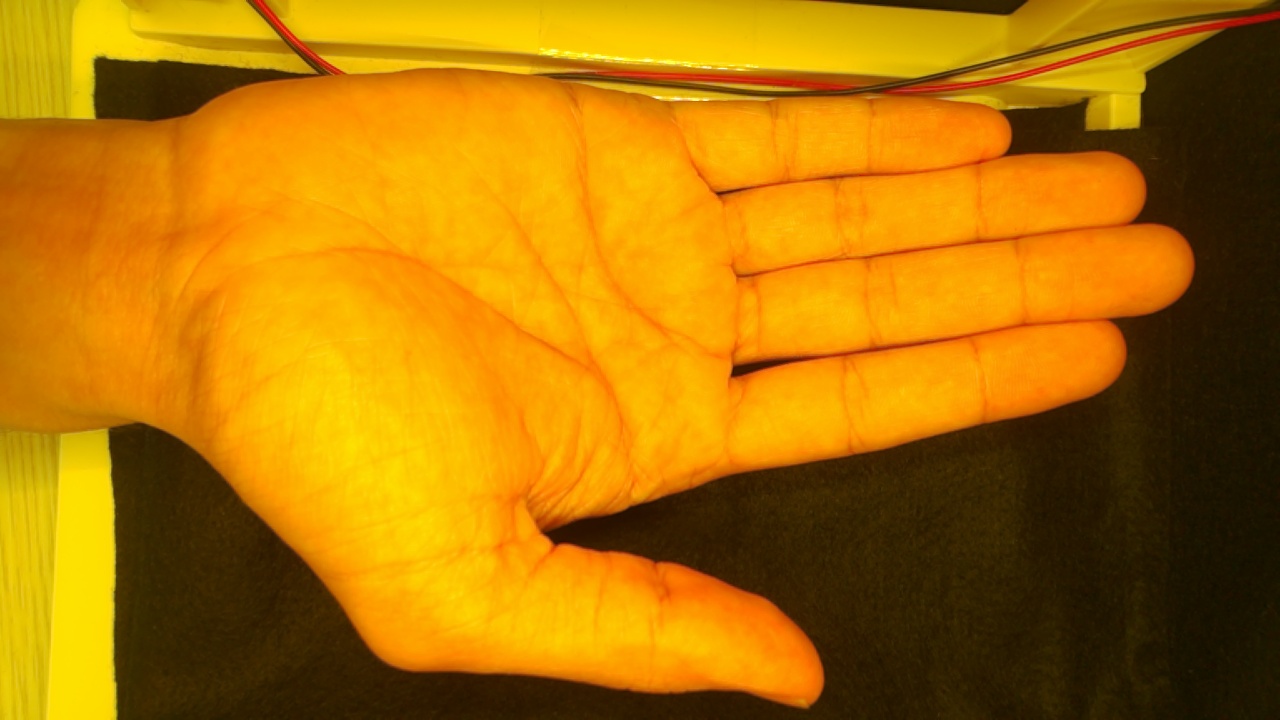}
                    \caption{Yellow}
                \end{subfigure}\hfill
                \begin{subfigure}{0.15\textwidth}
                    \includegraphics[width=\linewidth]{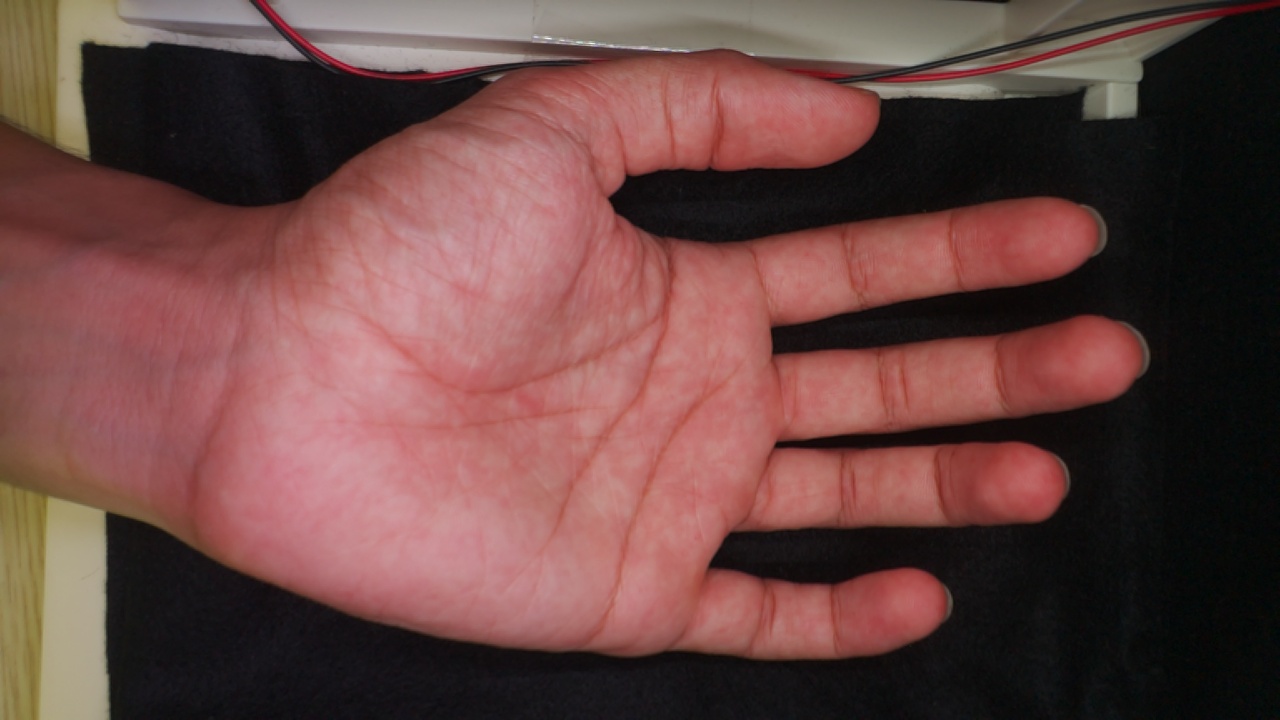}
                    \caption{White}
                \end{subfigure}\hfill
                \begin{subfigure}{0.15\textwidth}
                    \includegraphics[width=\linewidth]{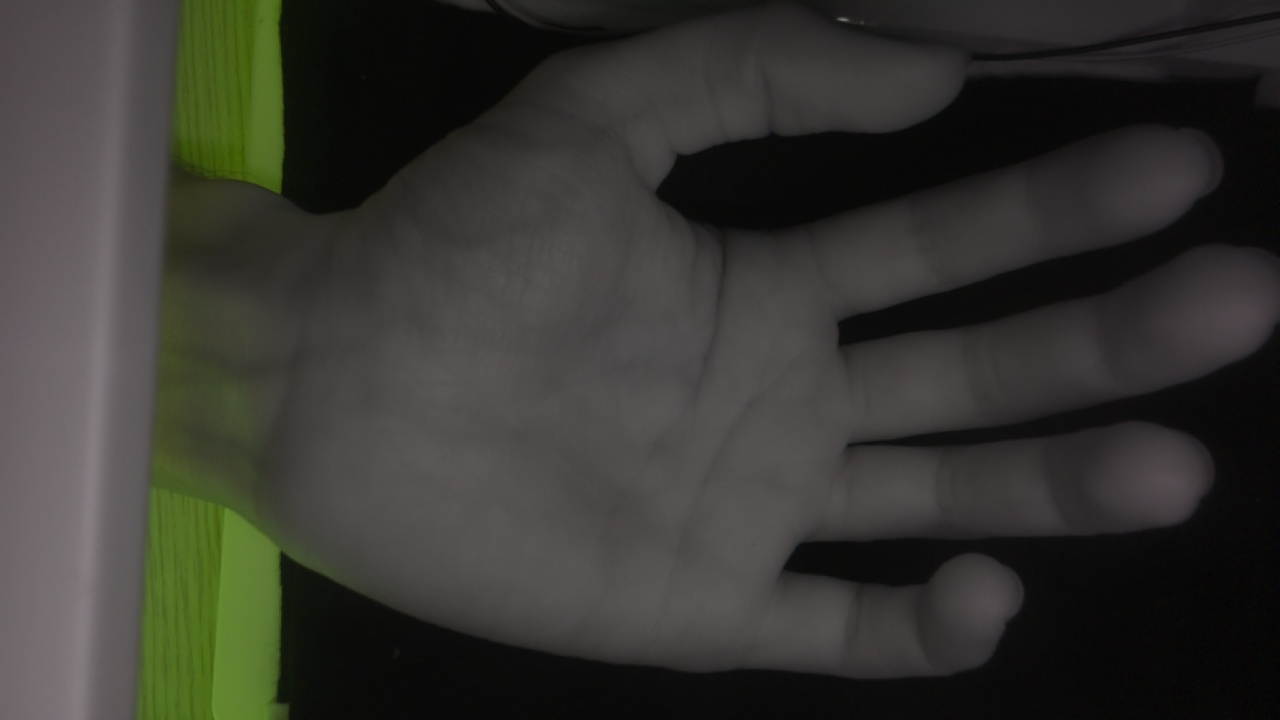}
                    \caption{IR}
                \end{subfigure}
            \end{minipage}
        };
        \draw [line width=1.5pt, rounded corners, gray!80] 
            ($(controlled.north west)$ ) rectangle ($(controlled.south east)$);
        \node[fill=white, text=black, font=\bfseries] at (controlled.north) {\small Controlled Multispectral Enrollment};

        \vspace{0.8cm} 

        \node (unconstrained) [below=0.5cm of controlled] {
            \begin{minipage}{\textwidth}
                \centering
                \begin{subfigure}{0.15\textwidth}
                    \includegraphics[width=\linewidth]{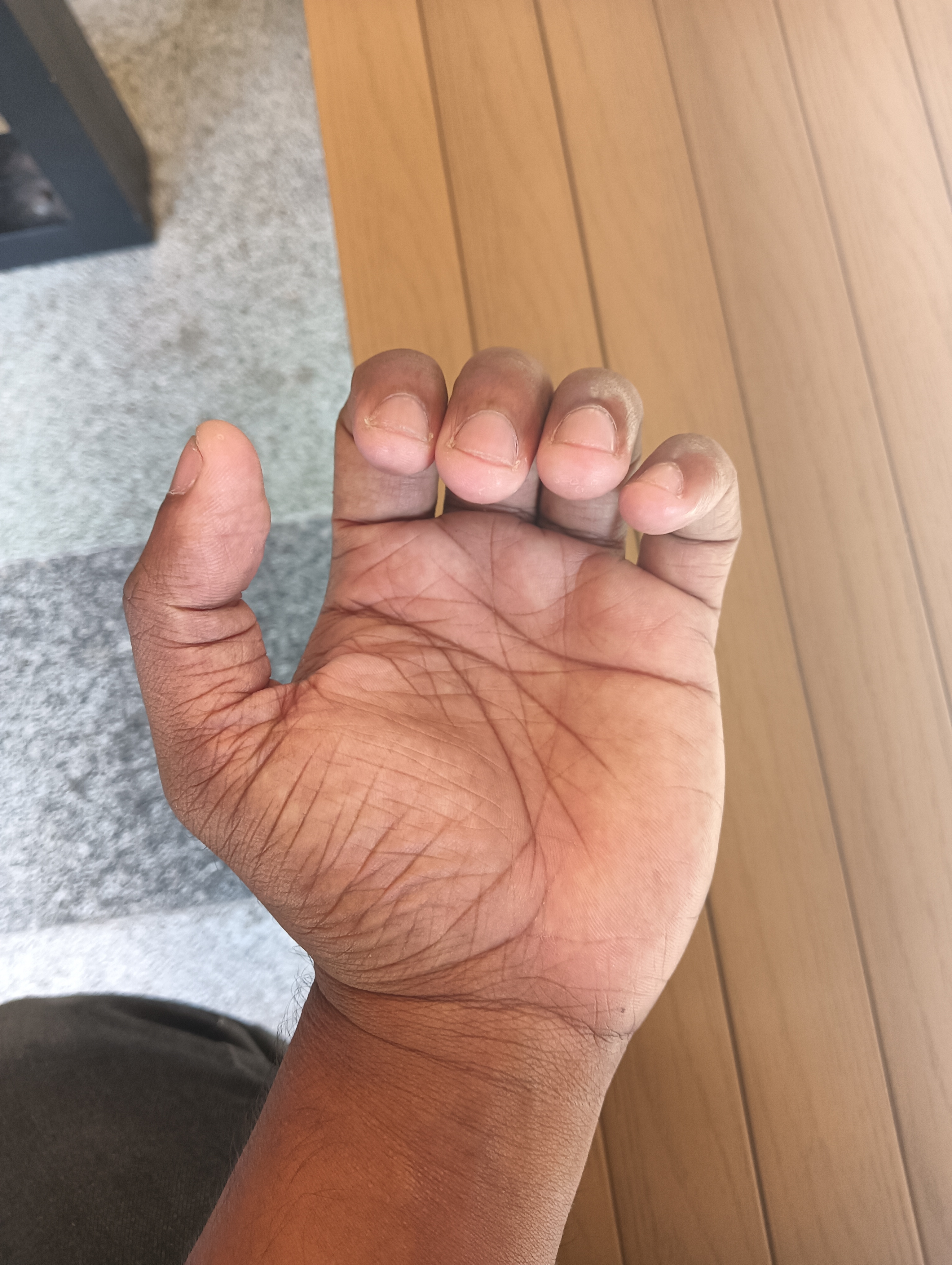}
                    \caption{Bent Fingers}
                \end{subfigure}\hfill
                \begin{subfigure}{0.15\textwidth}
                    \includegraphics[width=\linewidth]{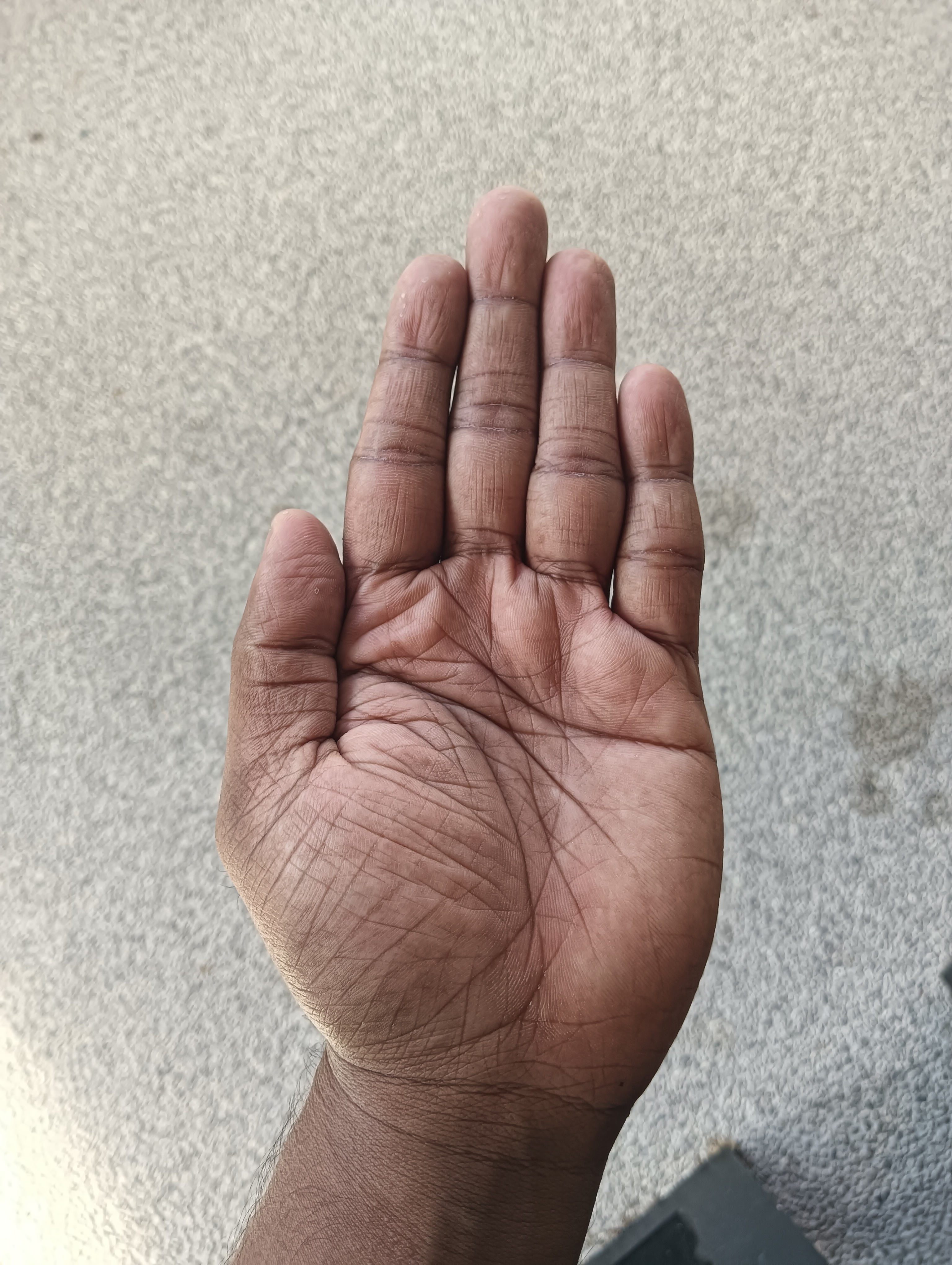}
                    \caption{Joined Fingers}
                \end{subfigure}\hfill
                \begin{subfigure}{0.15\textwidth}
                    \includegraphics[width=\linewidth]{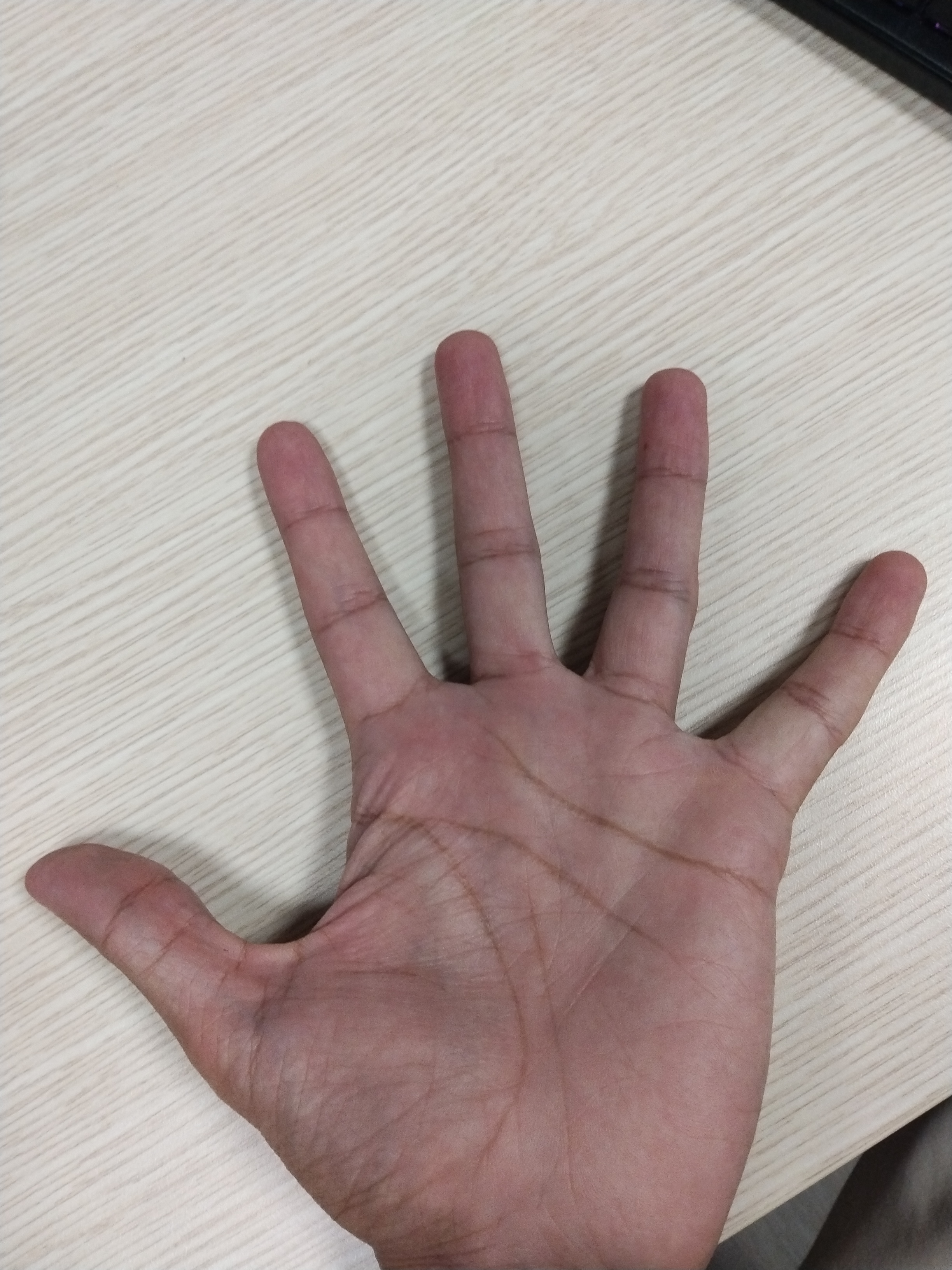}
                    \caption{Separate Fing.}
                \end{subfigure}\hfill
                \begin{subfigure}{0.15\textwidth}
                    \includegraphics[width=\linewidth]{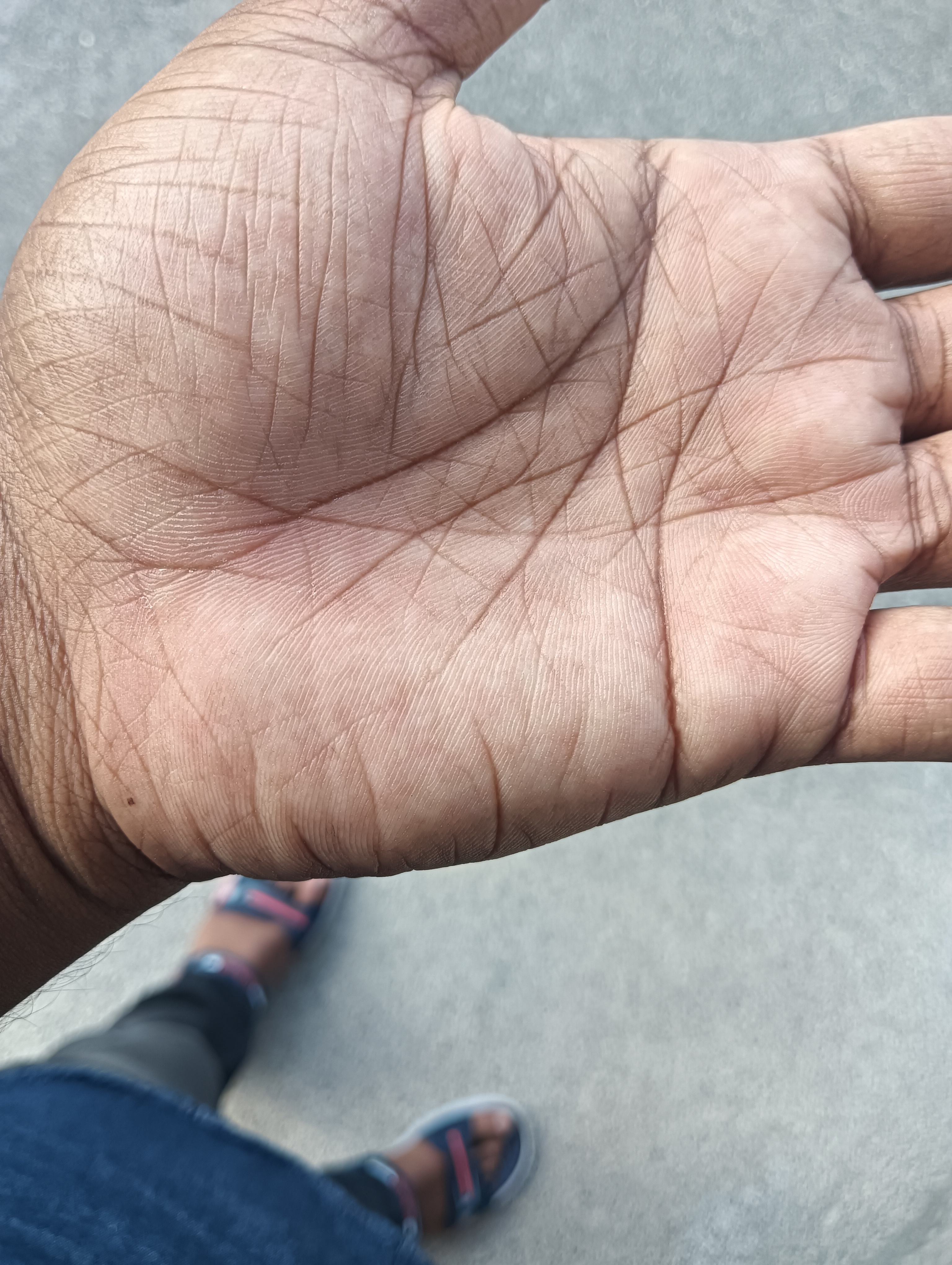}
                    \caption{Close}
                \end{subfigure}\hfill
                \begin{subfigure}{0.15\textwidth}
                    \includegraphics[width=\linewidth]{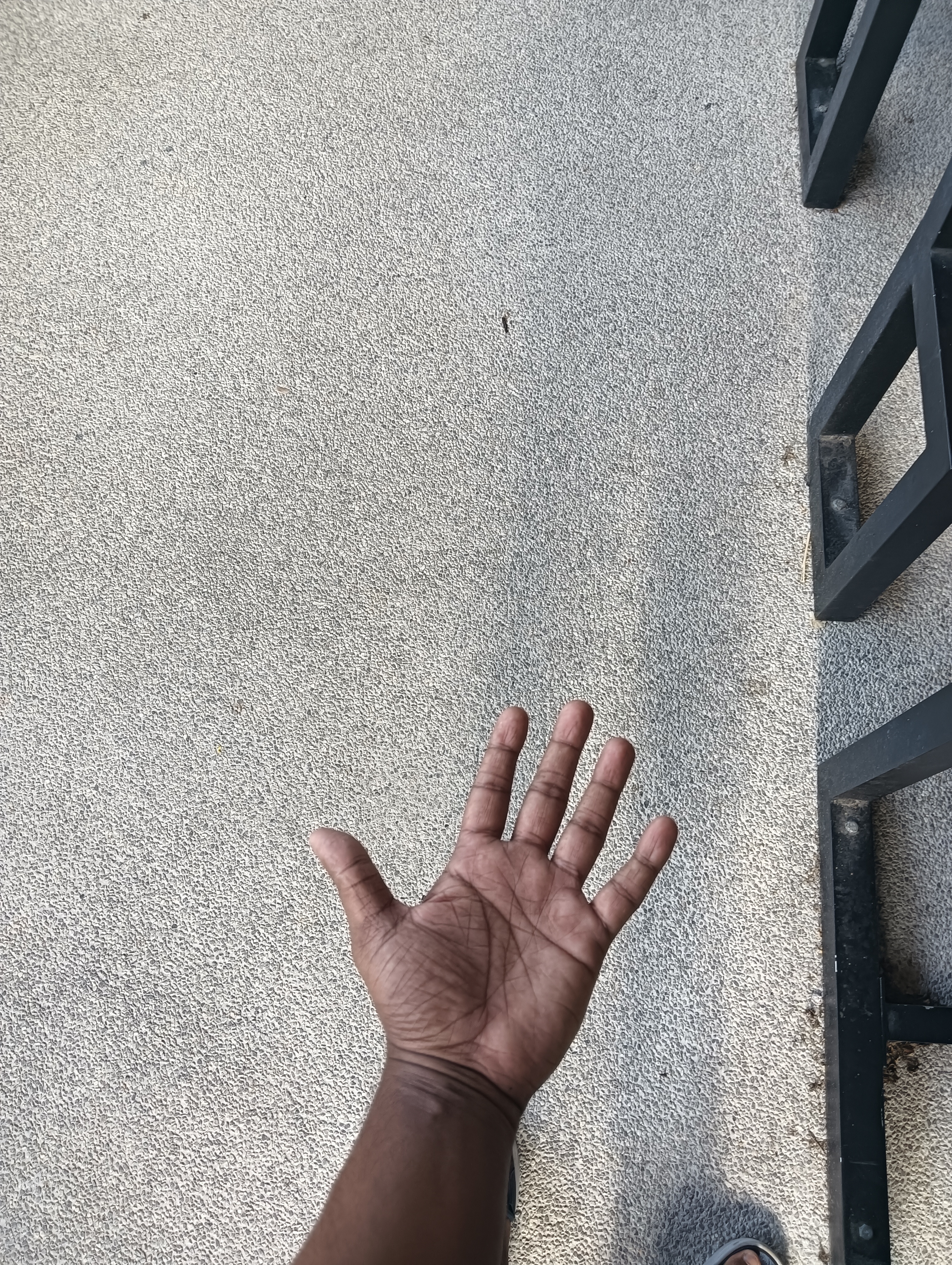}
                    \caption{Far}
                \end{subfigure}\hfill
                \begin{subfigure}{0.15\textwidth}
                    \includegraphics[width=\linewidth]{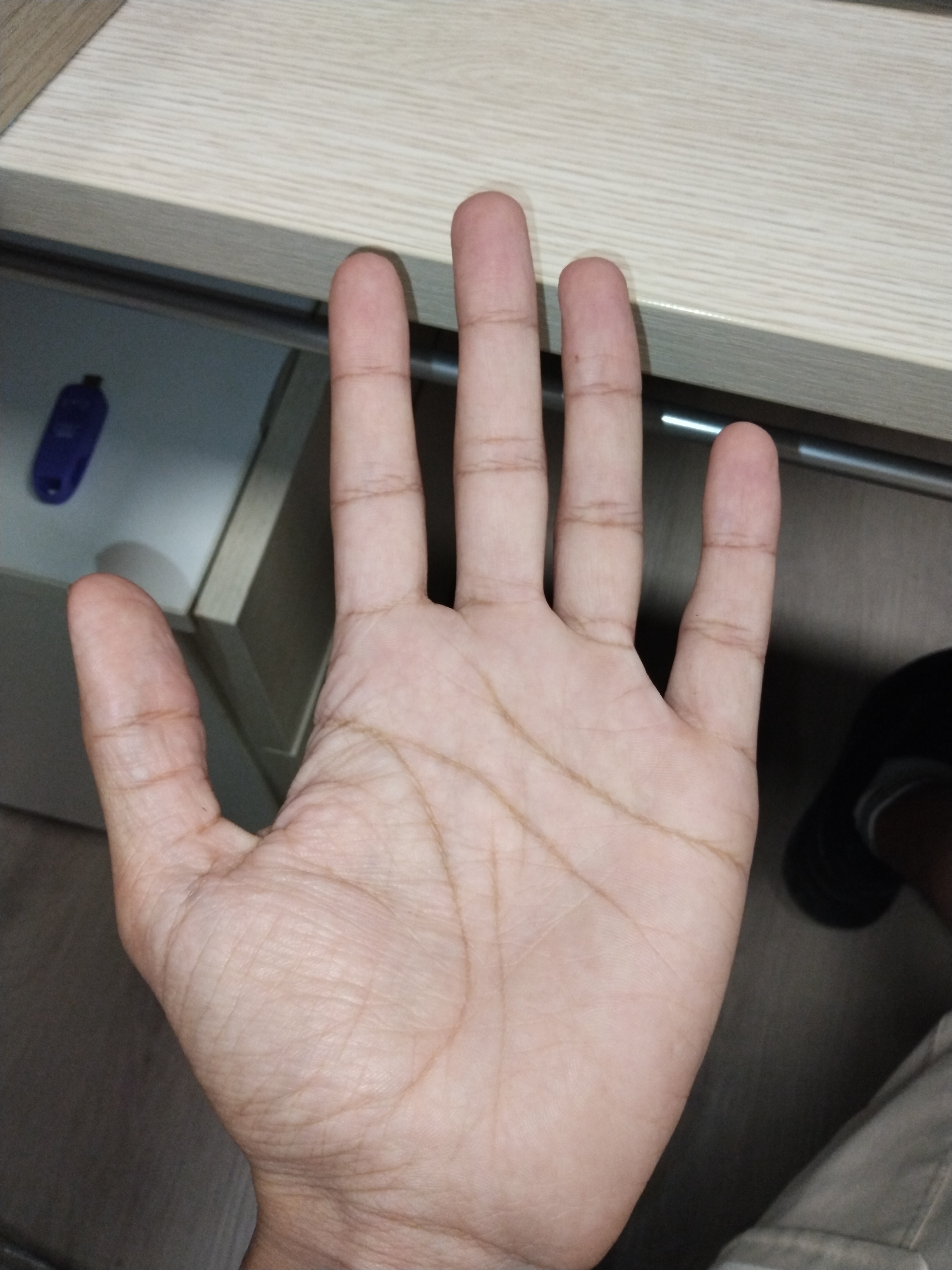}
                    \caption{Flash ON}
                \end{subfigure}

                \vspace{0.4cm}

                \begin{subfigure}{0.15\textwidth}
                    \includegraphics[width=\linewidth]{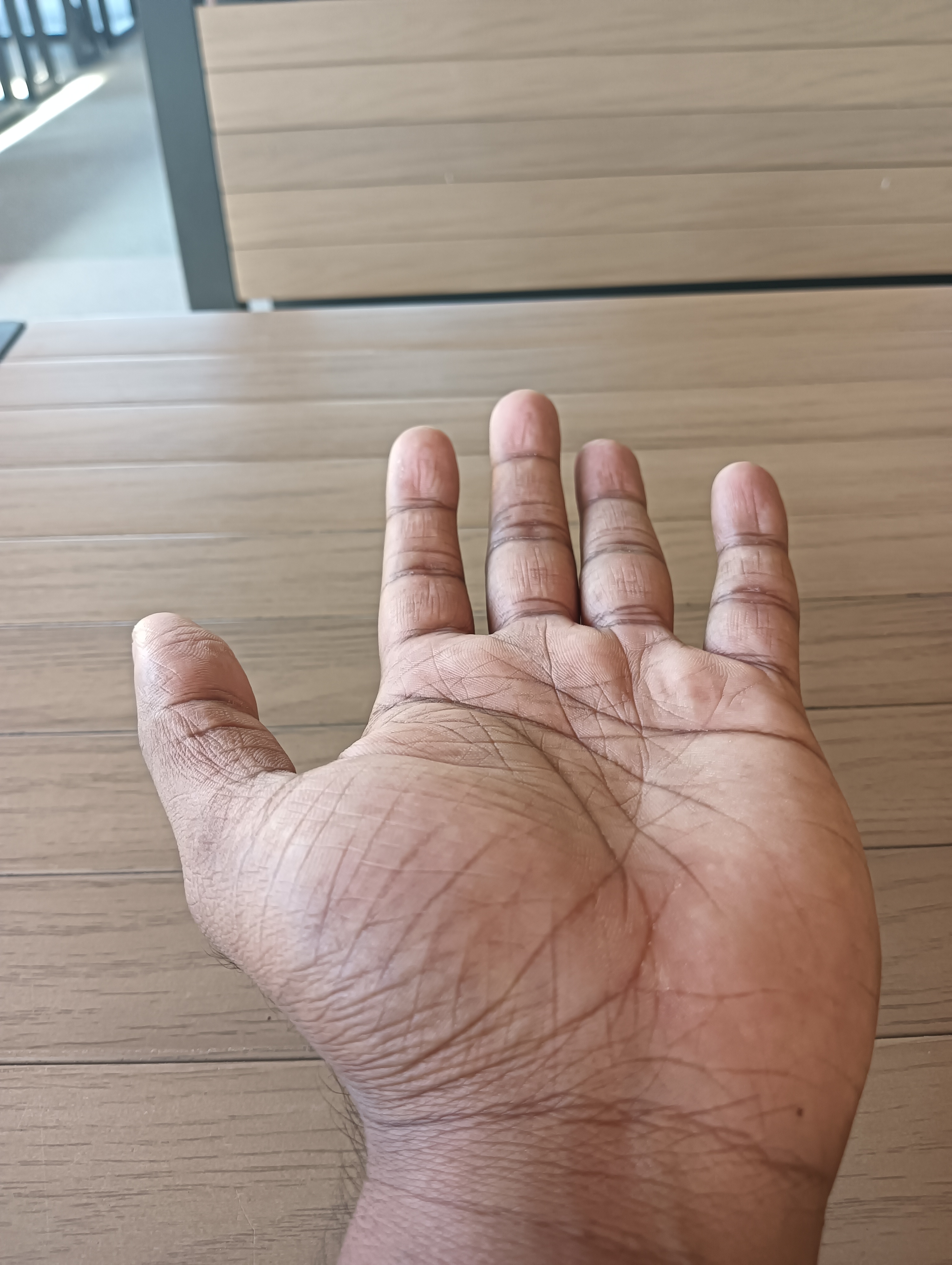}
                    \caption{Pitch}
                \end{subfigure}\hfill
                \begin{subfigure}{0.15\textwidth}
                    \includegraphics[width=\linewidth]{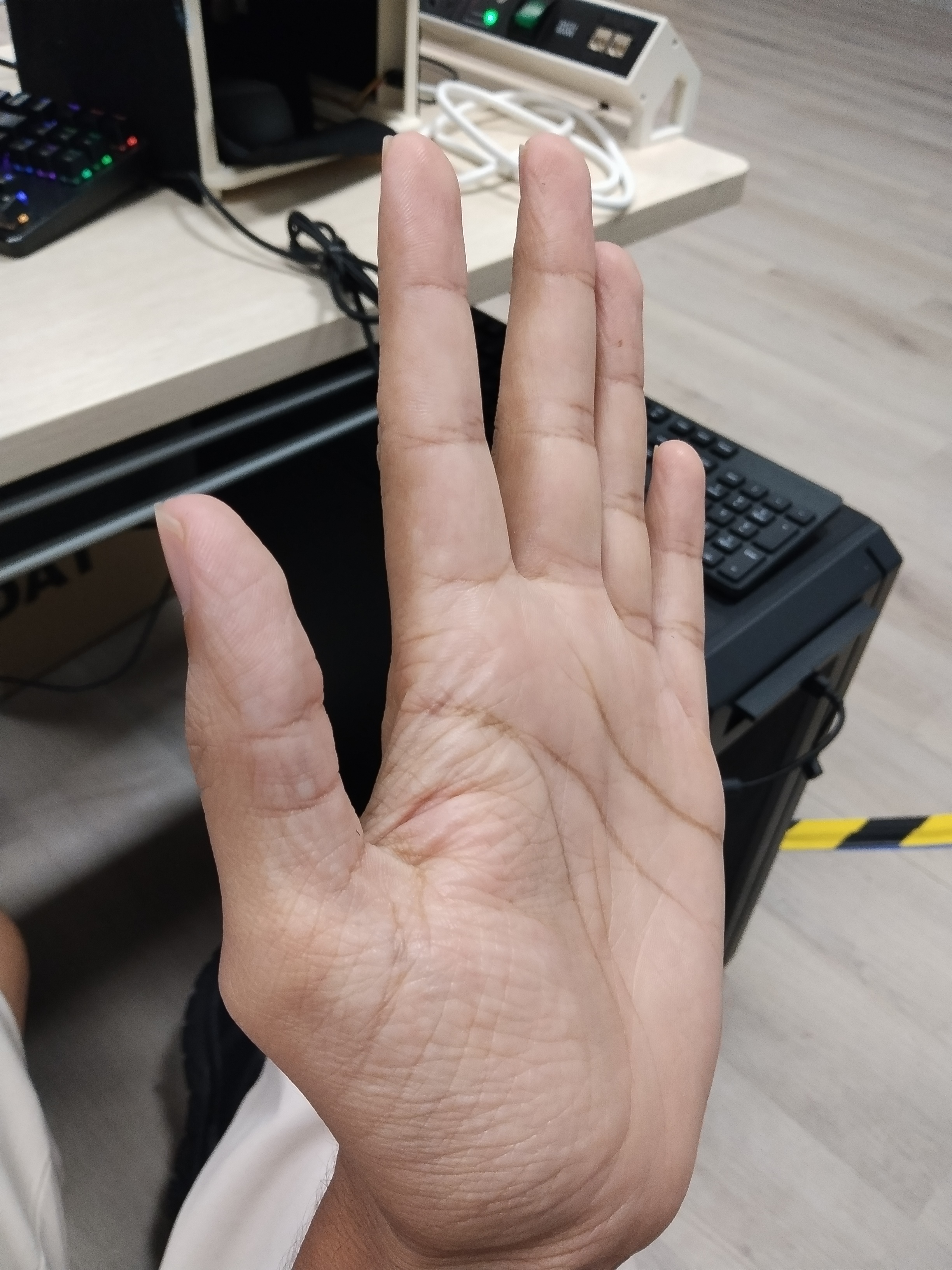}
                    \caption{Roll}
                \end{subfigure}\hfill
                \begin{subfigure}{0.15\textwidth}
                    \includegraphics[width=\linewidth]{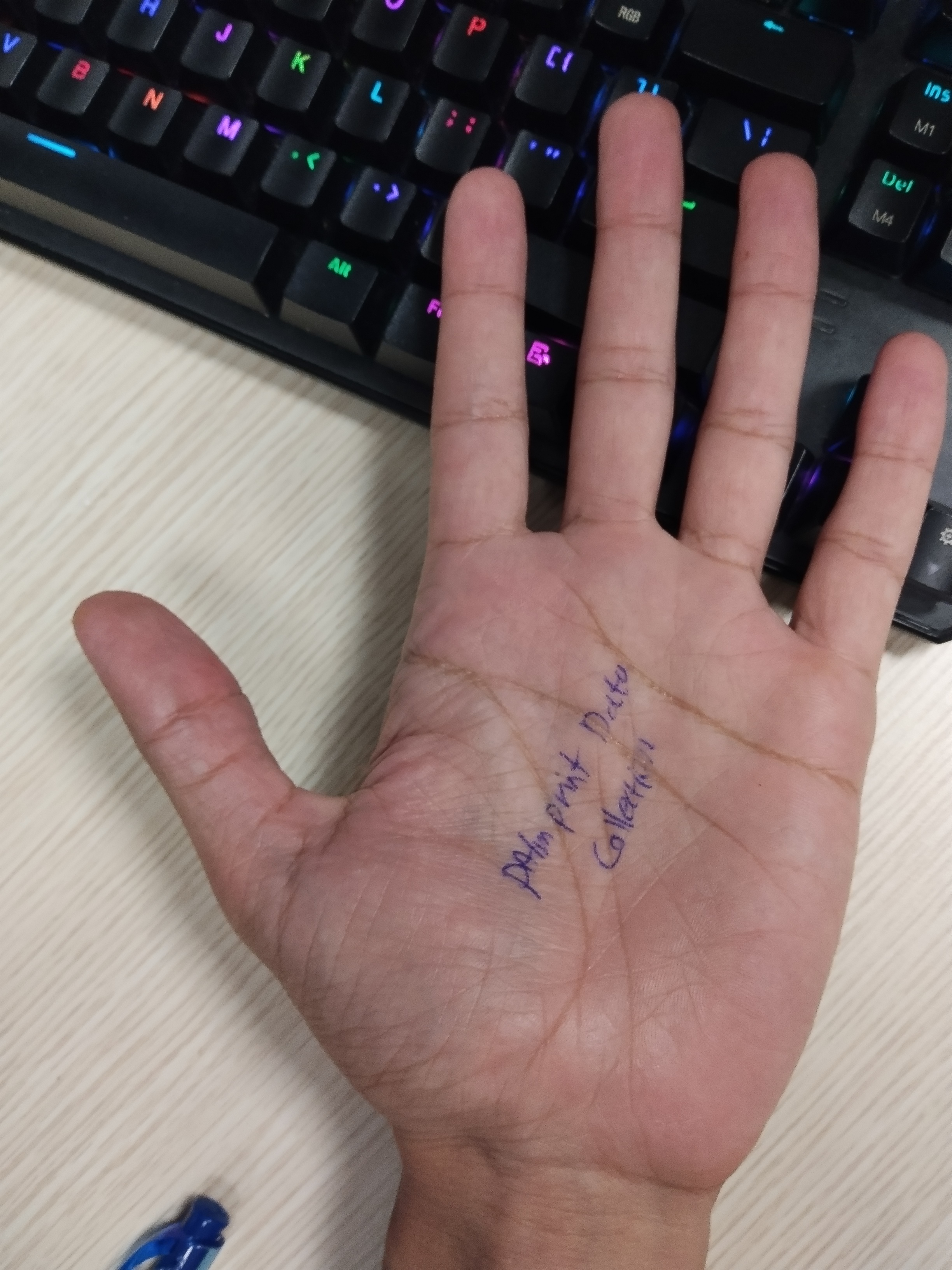}
                    \caption{Text}
                \end{subfigure}\hfill
                \begin{subfigure}{0.15\textwidth}
                    \includegraphics[width=\linewidth]{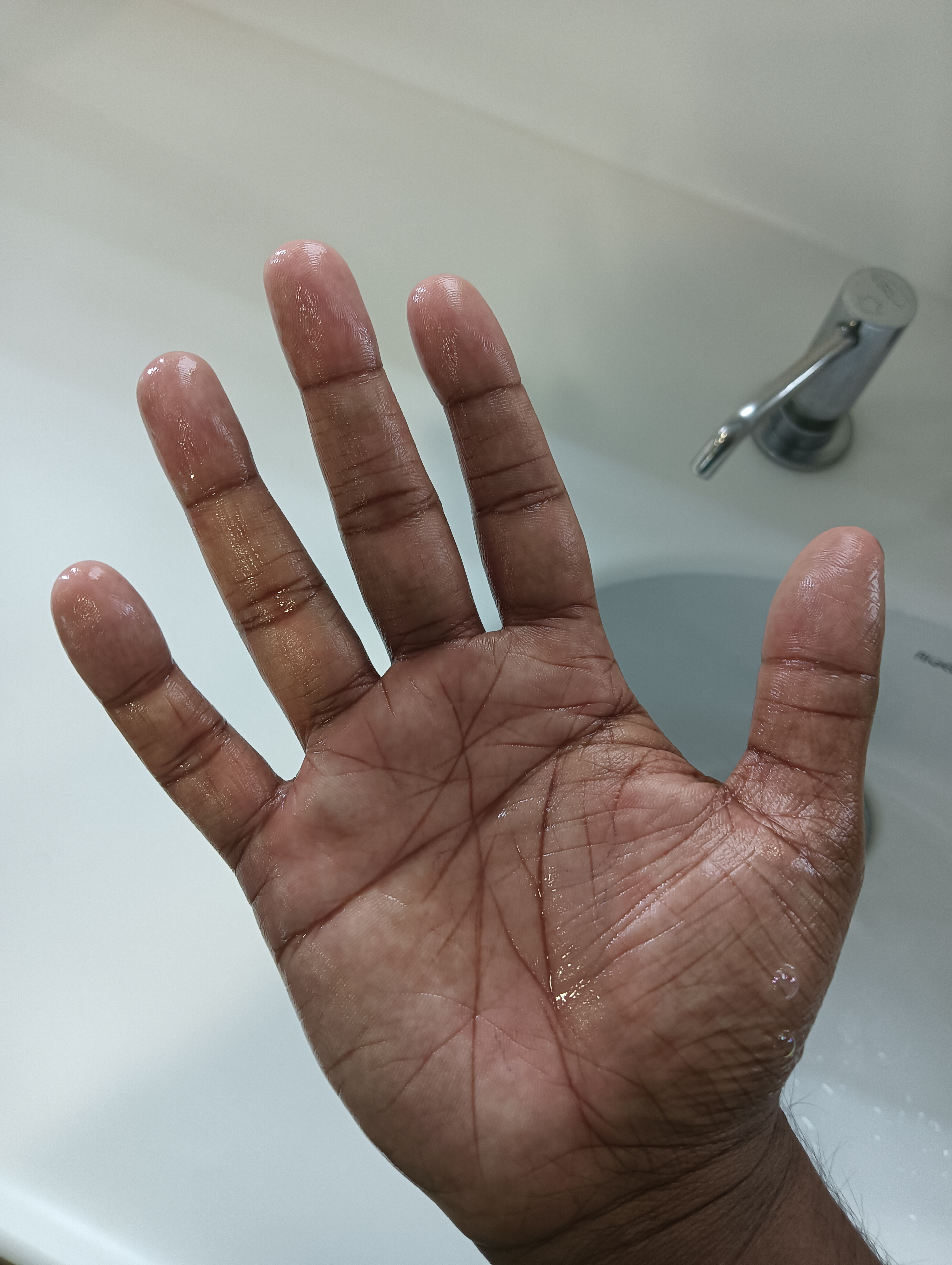}
                    \caption{Wet}
                \end{subfigure}\hfill
                \begin{subfigure}{0.15\textwidth}
                    \includegraphics[width=\linewidth]{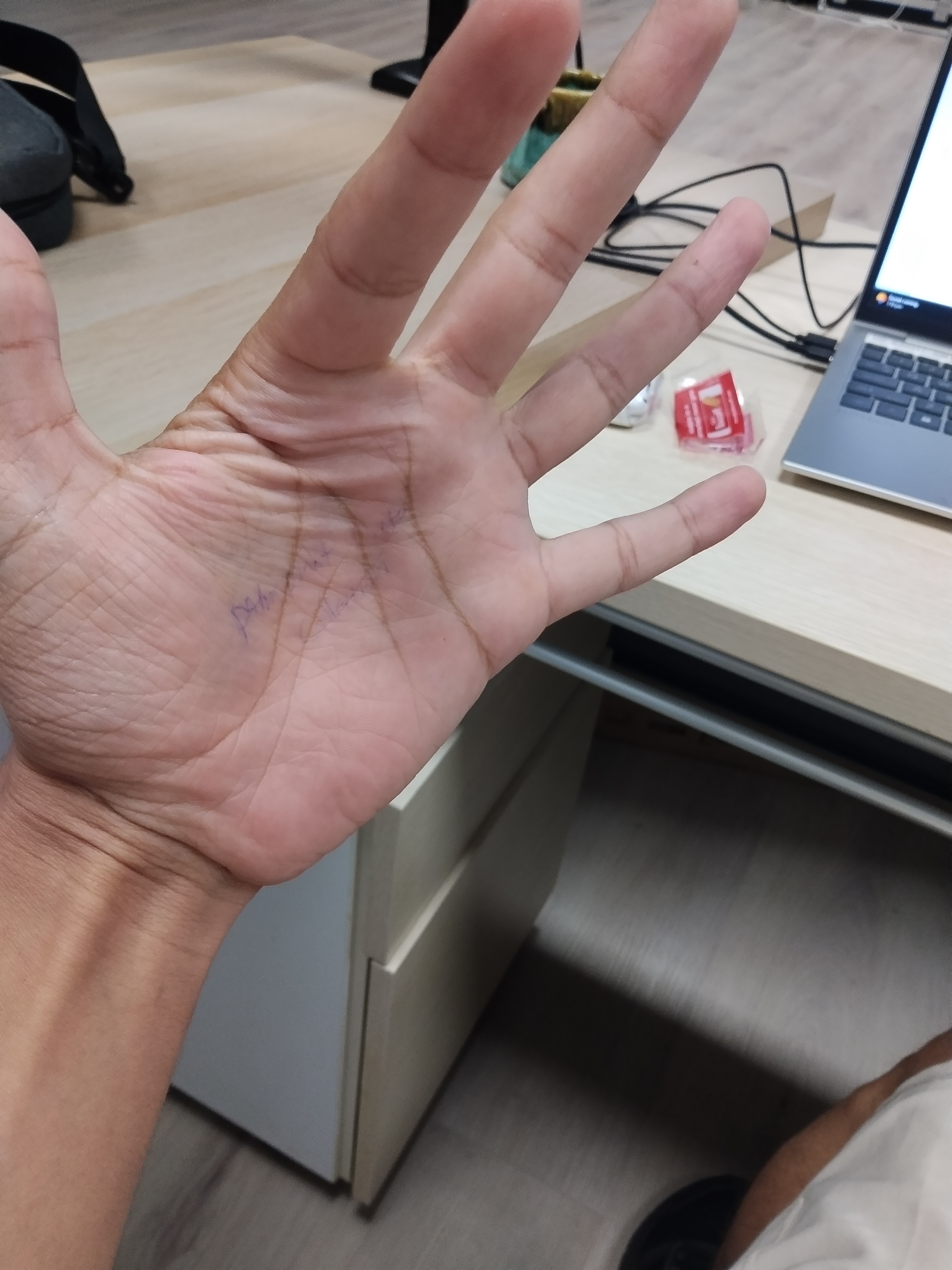}
                    \caption{Random 1}
                \end{subfigure}\hfill
                \begin{subfigure}{0.15\textwidth}
                    \includegraphics[width=\linewidth]{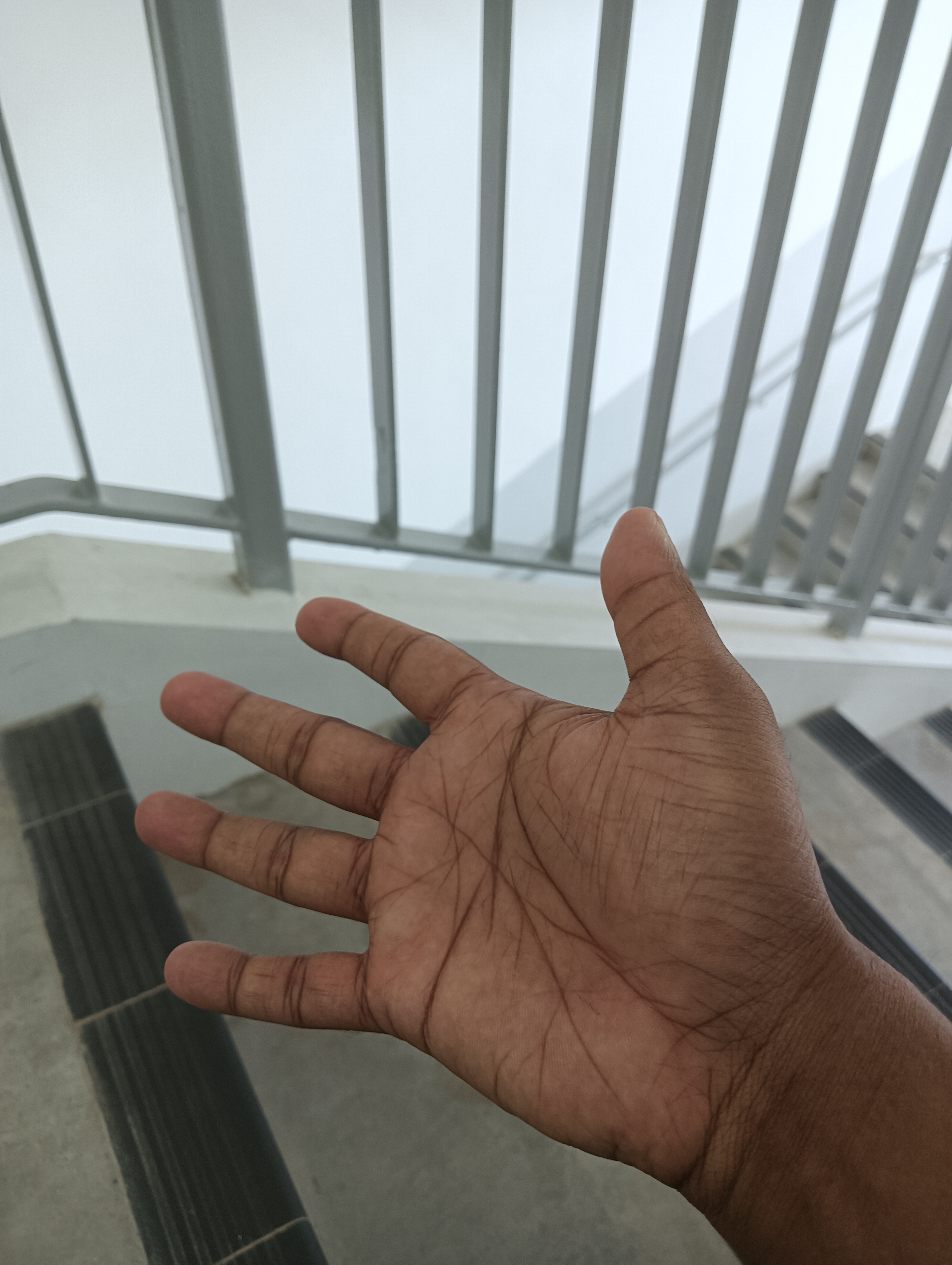}
                    \caption{Random 2}
                \end{subfigure}
            \end{minipage}
        };
        \draw [line width=1.5pt, rounded corners, blue!60] 
            ($(unconstrained.north west)$ ) rectangle ($(unconstrained.south east)$);
        \node[fill=white, text=blue!80, font=\bfseries] at (unconstrained.north) {\small Unconstrained Smartphone Authentication};

    \end{tikzpicture}
    
    \caption{Visual variations of the X-Palm dataset. 
    The top row (Gray Box) illustrates controlled multispectral enrollment templates captured across six spectral bands.
    The bottom two rows (Blue Box) demonstrate the unconstrained smartphone authentication domain, capturing simultaneous compound variability, including hand pose and perspective changes, distance shifts, and adverse surface and lighting conditions (e.g., moisture and occlusions).
    }
    \label{fig:orig_img}
\end{figure*}

\subsection{Data Collection Protocol}

Participants (n = 103) were recruited through email campaigns, and eligibility was restricted to adults aged 18 years or older. Prior to enrollment, each participant received information detailing the purpose of the research, the data acquisition procedures, the categories of biometric and demographic information to be collected, the intended use and public release of the de-identified dataset, and their right to withdraw at any time without penalty. Written informed consent was then obtained from every participant before any data was collected. Each participant received a small token of appreciation upon completion of the data collection session. All study procedures were reviewed and approved by the Institutional Review Board (IRB) of Singapore Institute of Technology under protocol number RECAS-0400.


For the Multispectral Palmprint Collection phase, participants were asked to position their hands within the scanner's designated capture area, directly facing the camera modules and programmable LED stripts. Although this acquisition environment was rigorously controlled to ensure spectral and illumination consistency, we deliberately introduced hand pose variations to capture structural diversity. Specifically, a set of 18 images (3 hand poses $\times$ 6 spectra) was acquired for each hand across three distinct hand poses: separated fingers, bent fingers, and joined fingers, as depicted in Fig.\ref{fig:orig_img}(a-f). Notably, the multispectral data is collected from 81 participants.

In the Smartphone Palmprint Collection phase, participants captured palm images entirely on their own smart devices, with supervised assistance limited only to those unable to perform the capture process themselves. Participants were guided through a Microsoft Form providing detailed instructions, starting with a demographic and hardware survey (including name, age, gender, ethnicity, and smartphone model). The collection protocol deliberately combined structured and random image acquisitions, comprising 10 structured and 5 random images for each hand. As shown in Fig.\ref{fig:orig_img}(g-p), structured images were carefully designed to ensure that a wide set of variations was captured for each participant, requiring specific changes in hand pose (separated, joined, bent), perspective (pitched, rolled, top-down), distance (close, far, normal), lighting (flash on/off), and surface condition (wrinkle, wet, text). Conversely, the random images are captured with arbitrary combinations of these factors, as shown in Fig.\ref{fig:orig_img}(q-r), to better simulate the unpredictability and challenges of practical, real-world deployment scenarios.

\subsection{Data Statistics}
Our dataset comprises 103 participants, 81 participants provided paired scanner and smartphone data (yielding 33 images per hand), while the remaining 22 participants provided exclusively smartphone data (15 images per hand). Participants (61 male, 59.22\%; 42 female, 40.78\%) aged 18--76 years, with the majority falling in the 18--25 (41.75\%) and 26--35 (25.24\%) age groups, while ensuring representation across adult age groups. Participants span 11 ethnic groups, with Chinese forming the largest group (59.22\%), followed by Indian (12.62\%), Iranian and Malay (7.77\% each), Indonesian (4.86\%), and Nepali (2.91\%), with five additional groups each contributing 0.97\%. This multi-ethnic composition reduces demographic bias and supports cross-population generalization. Smartphone captures cover 10 distinct device brands led by Apple (39.81\%), Samsung (16.50\%), and Google Pixel (15.53\%), with the remainder distributed across OPPO, Vivo, Xiaomi, and four other brands, collectively spanning over 80 distinct device models. This broad hardware diversity introduces realistic variation in sensor characteristics, image resolution, and capture quality. Full demographic breakdowns with visualizations are provided in Appendix~\ref{sec:demographics}.

\begin{figure*}[t]
    \centering
    \newlength{\annotatorheight}
    \setlength{\annotatorheight}{8cm}

    \begin{subfigure}[t]{0.70\textwidth}
        \vspace{0pt}
        \centering
        \begin{tikzpicture}
            \node[inner sep=0pt] (annotator) {
                \includegraphics[height=\annotatorheight, width=\linewidth, keepaspectratio]{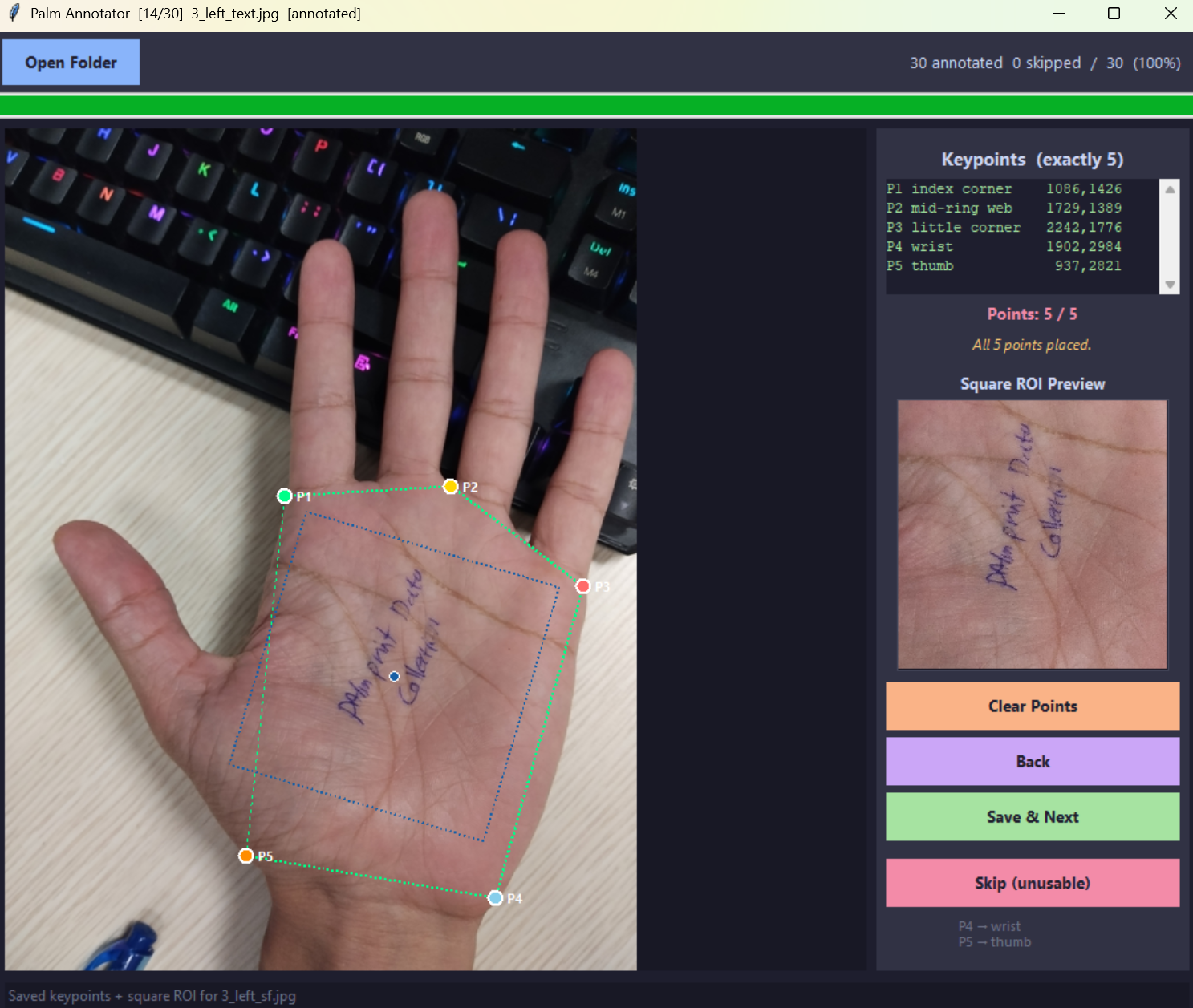}
            };
            
            \draw [line width=1.5pt, rounded corners, blue!60] 
                (annotator.north west) rectangle (annotator.south east);
            \node[fill=white, text=blue!80, font=\bfseries] at (annotator.north) {\small Data Annotation Interface};
        \end{tikzpicture}
        \label{fig:annotator_tool_left}
    \end{subfigure}
    \hfill
    \begin{subfigure}[t]{0.28\textwidth}
        \vspace{0pt}
        \centering
        \begin{tikzpicture}
            \node[inner sep=0pt, text width=\linewidth] (rois) {
                \begin{minipage}[t]{0.48\linewidth}
                    \includegraphics[height=0.24\annotatorheight, width=\linewidth, keepaspectratio]{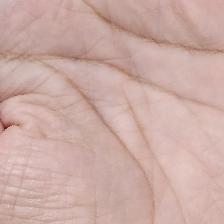}\\[2pt]
                    \includegraphics[height=0.24\annotatorheight, width=\linewidth, keepaspectratio]{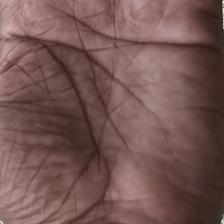}\\[2pt]
                    \includegraphics[height=0.24\annotatorheight, width=\linewidth, keepaspectratio]{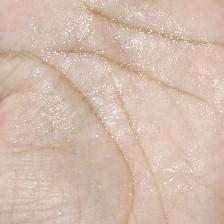}\\[2pt]
                    \includegraphics[height=0.24\annotatorheight, width=\linewidth, keepaspectratio]{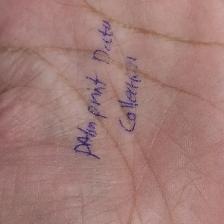}
                \end{minipage}%
                \hfill%
                \begin{minipage}[t]{0.48\linewidth}
                    \includegraphics[height=0.24\annotatorheight, width=\linewidth, keepaspectratio]{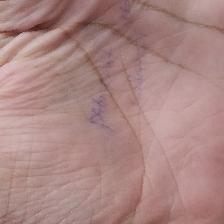}\\[2pt]
                    \includegraphics[height=0.24\annotatorheight, width=\linewidth, keepaspectratio]{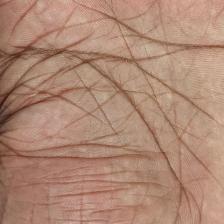}\\[2pt]
                    \includegraphics[height=0.24\annotatorheight, width=\linewidth, keepaspectratio]{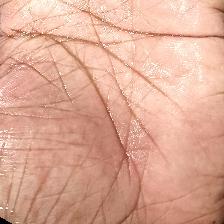}\\[2pt]
                    \includegraphics[height=0.24\annotatorheight, width=\linewidth, keepaspectratio]{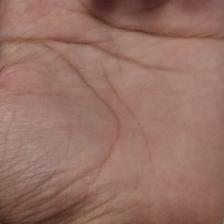}
                \end{minipage}
            };
            
            \draw [line width=1.5pt, rounded corners, red!60] 
                (rois.north west) rectangle (rois.south east);
            \node[fill=white, text=red!80, font=\bfseries] at (rois.north) {\small Extracted ROIs};
        \end{tikzpicture}
        \label{fig:sample_rois}
    \end{subfigure}
    \caption{The ROI is extracted using a semi-automated tool, where a human annotator manually marks five anatomical keypoints ($P_1$--$P_5$). Based on these manual anchors, the tool automatically extracts the ROI. Example extracted ROIs are shown on the right. Further details are provided in Appendix~\ref{roi}.}
    \label{fig:main_figure}
\end{figure*}

\subsection{Data Cleaning and Semi-Automated ROI Extraction}
Given the unconstrained and participant-driven nature of the smartphone collection, a rigorous curation phase is required to ensure dataset integrity.
Because participants have acquired images across highly variable environments and occasionally misinterpreted the capture instructions, a subset of the raw data suffers from artifacts, such as heavy motion blur, out-of-focus capture, poor lighting, or severe occlusion of the palm surface. To mitigate these issues, collected samples are subjected to a one-by-one manual quality inspection. 
To ensure a reliable ROI sample is extracted, we employ a semi-automated pipeline.
For each viable sample, a human annotator manually marked five specific anatomical keypoints. These keypoints are then utilized to establish a coordinate system, enabling automated extraction of the Region of Interest (ROI) of size $112 \times 112$, as shown in Fig.~\ref{fig:annotator_tool} (see Fig.\ref{fig:dataset_comparison_grid} in Appendix for visual comparison among ROIs of X-Palm and public datasets).


\section{Evaluations and Benchmarks}

\subsection{Evaluation Setup}
\label{evaluation_setup}
\begin{wraptable}{r}{0.48\textwidth}
    \vspace{-10pt}
    \centering
    \caption{Computational complexity.}
    \label{tab:complexity_comparison}
    \resizebox{0.48\textwidth}{!}{
    \begin{tabular}{lcc}
        \toprule
        \textbf{Model} & \textbf{Params} & \textbf{GFLOPs} \\
        \midrule
        CompNet~\cite{liang2021compnet} & 3.27M & 0.735 \\
        PPNet~\cite{liang2022ppnet} & 3.53M & 0.735 \\
        CCNet~\cite{yang2023ccnet} & 20.57M & 2.131 \\
        CO3Net~\cite{yang2023co3net} & 20.57M & 2.131 \\
        SF2Net~\cite{liu2025sf2net} & 13.08M & 2.655 \\
        PalmBridge~\cite{palmbridge} & 3.53M & 0.735 \\
        TSCAN~\cite{tscan} & 11.31M & 0.486 \\
        GIFT~\cite{gift} & 11.24M & 0.486 \\
        ConvNeXtV2-T~\cite{convnet} & 27.87M & 1.067 \\
        DINOv2-S/14~\cite{dino} & 22.06M & 1.398 \\
        ArcFace-iResNet100~\cite{arcface} & 65.12M & 12.098 \\
        MagFace-iResNet100~\cite{magface} & 65.16M & 12.117 \\
        \bottomrule
    \end{tabular}
    }
    \vspace{-10pt}
\end{wraptable}
\textbf{Baseline Methods. }
To establish a comprehensive benchmark, we evaluate our proposed dataset alongside other three public datasets (i.e., CASIA-MS, MPD-v2, and XJTU-UP) using state-of-the-art deep learning methods designed for palmprint recognition: CompNet \cite{liang2021compnet}, PPNet \cite{liang2022ppnet}, CCNet \cite{yang2023ccnet}, CO3Net \cite{yang2023co3net}, SF2Net \cite{liu2025sf2net}, and PalmBridge (with CompNet backbone) \cite{palmbridge}. In addition, we use domain adaptation and domain generalization methods in the palmprint literature: TSCAN \cite{tscan} and GIFT \cite{gift}. Furthermore, we employ pretrained models such as ConvNeXtV2-Tiny \cite{convnet}, DINOv2 ViT-S/14 \cite{dino}, ArcFace-iResNet100 \cite{arcface}, and MagFace-iResNet100 \cite{magface}. Table \ref{tab:complexity_comparison} outlines the computational complexity of these baselines, detailing both the total parameter count and the inference cost (in GFLOPs) for an input resolution of $112 \times 112$. 


\subsection{Evaluation Metrics and Task Formulation}

We evaluate baseline methods under three experimental settings: 
Cross-Dataset, Closed-Set Cross-Domain, and 
Open-Set Cross-Domain. Each setting targets a distinct 
generalization challenge and follows a gallery/probe evaluation protocol 
where the system matches probe images against a set of enrolled gallery 
templates using cosine similarity.

\textbf{Verification and Identification.} 
Verification (1:1 matching) confirms whether a probe image $x_p$ matches 
a claimed gallery template $x_g$. The system authenticates the user if 
the cosine similarity $S(f(x_p), f(x_g))$ exceeds a threshold $\tau$. 
We evaluate this using the Equal Error Rate (EER), which represents the 
operating point where the False Acceptance Rate (FAR) and False Rejection 
Rate (FRR) are equal:
\begin{equation}
    \text{EER} = \text{FAR}(\tau^*) = \text{FRR}(\tau^*)
\end{equation}
where $\tau^*$ is the threshold at which the two error rates intersect. 
A lower EER indicates a more robust system. Identification (1:N matching) 
determines the identity of an unknown probe $x_p$ by finding the template 
with the maximum cosine similarity within a gallery set $\mathcal{G}$:
\begin{equation}
    \hat{y} = \arg\max_{y_i \in \mathcal{Y}_\mathcal{G}} 
    S(f(x_p), f(x^i_g))
\end{equation}
where $\mathcal{Y}_\mathcal{G}$ represents the known identities in the 
gallery. We evaluate this using Rank-1 Accuracy, which is the percentage 
of probes correctly matched to their true identity's top gallery template.

\textbf{Cross-Dataset Setting.} 
Models are trained on a source dataset and evaluated on a different target 
dataset. To ensure a fair comparison, we standardize the evaluation by 
using an equal number of identities and samples per identity across all 
datasets. The samples of each identity in the test dataset are partitioned 
into gallery and probe sets using a 50\% split.

\textbf{Closed-Set Cross-Domain Setting.} 
Models are trained on specific acquisition domains within X-Palm (e.g., 
the Smartphone collection) and tested on held-out domains (e.g., the 
Multispectral Scanner collection). Critically, the identities are shared 
between train and test sets ($\mathcal{Y}_{train} = \mathcal{Y}_{test}$), 
isolating domain shift as the sole source of difficulty. The samples of 
each identity in the test domains are partitioned into gallery and probe 
sets using a 50\% split.

\textbf{Open-Set Cross-Domain Setting.} 
This setting extends the closed-set protocol to a more practical and 
challenging scenario where the identities in the test set are entirely 
disjoint from those in the training set 
($\mathcal{Y}_{train} \cap \mathcal{Y}_{test} = \emptyset$). Models are 
trained on specific domains of the training identities and evaluated on 
different domains of unseen test identities, assessing their capacity to 
generalize across simultaneous domain shifts and unseen identities. 
20\% of identities are assigned as test identities, with their samples 
partitioned into gallery and probe sets using a 50\% split.

\subsection{Experimental Results}
\label{results}
We benchmark the baseline methods under the three settings defined above. 
In Tables~\ref{tab:merged_results_models}--\ref{tab:eer_openset_cross_domain}, reported results are mean values across three independent runs, with detailed statistical significance analyses provided in Appendix~\ref{stat_sig}. 
In Tables~\ref{tab:rank1_cdcs_evaluation}--\ref{tab:eer_openset_cross_domain}, train domains in each row comprise all domains in the dataset excluding the test domain. RND, SF, JF, BF, and FO stand for random, separated fingers, joined fingers, bent fingers, and flash on, respectively. 
The best result in each row is highlighted in \textbf{bold} and the  second-best is \underline{underlined}. 
   
\begin{table*}[t]
  \centering
  \caption{Cross-Dataset Evaluation with EER (\%,$\downarrow$) and Rank-1 Accuracy (\%,$\uparrow$) across Different Models}
  \label{tab:merged_results_models}
  \resizebox{\textwidth}{!}{
  \begin{tabular}{l l cc cc cc cc | cc}
    \toprule
    \multirow{2}{*}{\textbf{Model}} & \multirow{2}{*}{\textbf{Train \textbackslash{} Test}} & \multicolumn{2}{c}{\textbf{X-Palm}} & \multicolumn{2}{c}{\textbf{CASIA-MS}} & \multicolumn{2}{c}{\textbf{MPD-v2}} & \multicolumn{2}{c}{\textbf{XJTU}} & \multicolumn{2}{c}{\textbf{Average}} \\
    \cmidrule(lr){3-4} \cmidrule(lr){5-6} \cmidrule(lr){7-8} \cmidrule(lr){9-10} \cmidrule(lr){11-12}
    & & \textbf{EER} & \textbf{Rank-1} & \textbf{EER} & \textbf{Rank-1} & \textbf{EER} & \textbf{Rank-1} & \textbf{EER} & \textbf{Rank-1} & \textbf{EER} & \textbf{Rank-1} \\
    \midrule
    \multirow{4}{*}{\textbf{CompNet}~\cite{liang2021compnet}}
    & X-Palm   & 28.98 & 79.70 & 13.14 & 96.56 & 22.29 & 86.05 & 6.06  & 99.46  & \textbf{17.62} & \textbf{90.44} \\
    & CASIA-MS & 36.97 & 66.30 & 9.12  & 100.00 & 22.08 & 87.12 & 4.27  & 99.91  & 18.11 & 88.33 \\
    & MPD-v2   & 35.91 & 66.92 & 14.27 & 99.15  & 17.04 & 91.95 & 4.93  & 99.51  & \underline{18.04} & \underline{89.38} \\
    & XJTU     & 36.09 & 67.04 & 13.30 & 100.00 & 20.26 & 88.39 & 3.76  & 100.00 & 18.35 & 88.86 \\
    \midrule
    \multirow{4}{*}{\textbf{PPNet}~\cite{liang2022ppnet}}
    & X-Palm   & 33.60 & 68.63 & 23.77 & 88.19 & 29.18 & 76.10 & 16.31 & 96.72  & 25.72 & \textbf{82.41} \\
    & CASIA-MS & 39.40 & 53.15 & 14.74 & 98.68 & 30.76 & 77.09 & 12.66 & 98.50  & \textbf{24.39} & \underline{81.86} \\
    & MPD-v2   & 38.34 & 52.07 & 24.75 & 91.26 & 24.12 & 85.90 & 13.57 & 98.06  & \underline{25.20} & 81.82 \\
    & XJTU     & 41.39 & 36.78 & 22.06 & 96.38 & 29.42 & 77.33 & 12.57 & 99.65  & 26.36 & 77.54 \\
    \midrule
    \multirow{4}{*}{\textbf{CCNet}~\cite{yang2023ccnet}}
    & X-Palm   & 26.99 & 77.82 & 15.43 & 93.48 & 23.22 & 82.23 & 8.17  & 98.85  & \textbf{18.45} & \textbf{88.10} \\
    & CASIA-MS & 37.34 & 61.06 & 9.80  & 99.88 & 23.19 & 85.41 & 6.24  & 99.63  & 19.14 & 86.50 \\
    & MPD-v2   & 35.76 & 63.19 & 15.53 & 98.14 & 17.71 & 88.52 & 6.29  & 99.15  & \underline{18.82} & 87.25 \\
    & XJTU     & 36.72 & 64.85 & 13.38 & 100.00 & 20.57 & 87.43 & 4.66  & 100.00 & 18.83 & \underline{88.07} \\
    \midrule
    \multirow{4}{*}{\textbf{CO3Net}~\cite{yang2023co3net}}
    & X-Palm   & 27.75 & 75.10 & 16.77 & 91.77 & 25.54 & 78.76 & 9.38  & 98.77  & 19.86 & \underline{86.10} \\
    & CASIA-MS & 37.63 & 58.24 & 9.26  & 100.00 & 24.71 & 82.51 & 7.24  & 99.27  & \underline{19.71} & 85.00 \\
    & MPD-v2   & 36.07 & 58.70 & 15.94 & 97.56 & 19.21 & 89.53 & 6.63  & 99.15  & \textbf{19.46} & \textbf{86.24} \\
    & XJTU     & 37.04 & 58.94 & 14.49 & 100.00 & 21.39 & 84.93 & 6.51  & 99.65  & 19.86 & 85.88 \\
    \midrule
    \multirow{4}{*}{\textbf{SF2Net}~\cite{liu2025sf2net}}
    & X-Palm   & 28.70 & 79.73 & 13.48 & 96.41 & 22.10 & 83.81 & 6.04  & 99.43  & \textbf{17.58} & \textbf{89.84} \\
    & CASIA-MS & 38.32 & 71.61 & 9.73  & 99.94 & 22.85 & 84.96 & 5.01  & 99.57  & 18.98 & 89.02 \\
    & MPD-v2   & 37.12 & 71.32 & 15.04 & 98.95 & 17.19 & 89.32 & 4.46  & 99.65  & \underline{18.45} & \underline{89.81} \\
    & XJTU     & 36.87 & 72.31 & 13.84 & 100.00 & 20.72 & 86.56 & 4.93  & 100.00 & 19.09 & 89.72 \\
    \midrule
    \multirow{4}{*}{\textbf{PalmBridge}~\cite{palmbridge}}
    & X-Palm   & 23.31 & 79.49 & 16.28 & 89.75 & 19.58 & 89.15 & 7.98  & 98.25  & \textbf{16.79} & \textbf{89.16} \\
    & CASIA-MS & 35.06 & 63.00 & 8.75  & 99.28 & 21.07 & 87.52 & 7.97  & 98.90  & \underline{18.21} & \underline{87.18} \\
    & MPD-v2   & 33.23 & 59.22 & 20.06 & 82.31 & 14.89 & 91.75 & 8.12  & 97.02  & 19.08 & 82.58 \\
    & XJTU     & 33.82 & 58.18 & 18.14 & 90.03 & 19.50 & 86.44 & 5.38  & 99.65  & 19.21 & 83.58 \\
    \midrule
    \multirow{4}{*}{\textbf{ConvNeXt}~\cite{convnet}}
    & X-Palm   & 22.55 & 67.52 & 26.26 & 58.72 & 22.54 & 66.58 & 16.06 & 85.14  & \textbf{21.85} & \textbf{69.49} \\
    & CASIA-MS & 37.07 & 26.21 & 15.26 & 87.17 & 25.29 & 66.31 & 17.33 & 83.97  & \underline{23.74} & \underline{65.91} \\
    & MPD-v2   & 39.19 & 27.35 & 32.21 & 62.13 & 16.25 & 86.58 & 18.55 & 81.06  & 26.55 & 64.28 \\
    & XJTU     & 36.96 & 25.82 & 30.47 & 67.16 & 22.97 & 70.78 & 14.23 & 91.27  & 26.16 & 63.76 \\
    \midrule
    \multirow{4}{*}{\textbf{DINOv2}~\cite{dino}}
    & X-Palm   & 20.74 & 76.08 & 28.11 & 55.41 & 20.90 & 70.52 & 18.30 & 82.24  & \textbf{22.01} & \textbf{71.06} \\
    & CASIA-MS & 35.62 & 35.11 & 15.55 & 88.17 & 24.23 & 69.57 & 21.23 & 79.79  & 24.16 & \underline{68.16} \\
    & MPD-v2   & 34.28 & 33.58 & 32.24 & 55.98 & 12.54 & 92.22 & 18.20 & 83.83  & 24.32 & 66.40 \\
    & XJTU     & 33.09 & 36.38 & 29.76 & 55.41 & 21.97 & 68.67 & 11.59 & 93.17  & \underline{24.10} & 63.41 \\
    \midrule
    \multirow{4}{*}{\textbf{ArcFace}~\cite{arcface}}
    & X-Palm   & 26.54 & 82.52 & 19.03 & 96.59 & 25.55 & 85.10 & 12.79 & 98.28  & \textbf{20.98} & \textbf{90.62} \\
    & CASIA-MS & 48.09 & 26.15 & 10.21 & 99.81 & 34.88 & 72.53 & 42.34 & 62.40  & 33.88 & 65.22 \\
    & MPD-v2   & 44.01 & 35.19 & 22.06 & 95.96 & 17.17 & 91.35 & 46.05 & 65.66  & 32.32 & 72.04 \\
    & XJTU     & 39.38 & 60.07 & 19.96 & 97.11 & 23.16 & 86.01 & 5.86  & 99.86  & \underline{22.09} & \underline{85.76} \\
    \midrule
    \multirow{4}{*}{\textbf{MagFace}~\cite{magface}}
    & X-Palm   & 25.39 & 76.05 & 19.12 & 95.58 & 28.24 & 82.08 & 14.20 & 97.20  & \textbf{21.74} & \textbf{87.73} \\
    & CASIA-MS & 46.60 & 25.23 & 9.72  & 99.81 & 38.82 & 64.13 & 46.48 & 52.03  & 35.40 & 60.30 \\
    & MPD-v2   & 40.26 & 40.44 & 23.47 & 92.74 & 21.17 & 88.53 & 27.90 & 90.02  & 28.20 & 77.93 \\
    & XJTU     & 37.68 & 46.33 & 17.90 & 96.61 & 25.28 & 84.46 & 8.33  & 99.86  & \underline{22.30} & \underline{81.82} \\
    \bottomrule
  \end{tabular}
  }
\end{table*}


\textbf{Cross-Dataset Evaluation}
As shown in 
Table~\ref{tab:merged_results_models}, when X-Palm is used for training, performance on other datasets remains highly comparable to models trained on those datasets directly. Conversely, when X-Palm is used for evaluation, models trained on other datasets experience significant performance degradation, underscoring the dataset's inherent difficulty. 
Models trained on X-Palm mostly yield superior average performance, as shown in the last two columns of Table~\ref{tab:merged_results_models}. 
This demonstrates that exposure to compound variations and diverse capture challenges forces models to learn more robust features. Consequently, X-Palm serves both as an effective training source and as a rigorous benchmark for evaluating true cross-domain generalization.
palmprint-specific models mostly outperform general-purpose and face-pretrained models despite having significantly lower computational cost. Among larger models, face-pretrained ArcFace and MagFace outperform general object classification models (ConvNeXt and DINOv2), which is expected as both face and palmprint recognition share similar discriminative objectives, learning fine-grained identity-specific features from biometric surfaces, making face-domain pretraining a more 
compatible initialization than ImageNet features trained for semantic object categorization.
(see Appendix~\ref{sec:appendix_results} for additional results).


\textbf{Closed-Set Cross-Domain Evaluation}
The first two rows of Tables~\ref{tab:rank1_cdcs_evaluation} 
and~\ref{tab:eer_cdcs_evaluation} show the relatively poor performance of all baselines in cross-setting authentication scenarios, specifically, when models are trained on smartphone data and evaluated on scanner data, or vice versa. The remaining rows evaluate baselines under different in-the-wild variations, demonstrating that \textit{Roll}, \textit{Pitch}, \textit{Wet}, \textit{Text}, and \textit{Random} domains present the most challenging scenarios for authentication. As illustrated in Fig.~\ref{fig:orig_img} and Fig.~\ref{fig:dataset_comparison_grid}, these conditions introduce severe intra-class variability: rolled and pitched captures drastically alter the visible palm geometry and partially occlude discriminative line patterns, wet surfaces blur and distort fine-grained creases and wrinkles, handwritten text directly occludes the palm surface, and random captures combine simultaneous variations in an unpredictable manner.

\begin{table*}[h]
    \centering
    \caption{Closed-Set Cross-Domain Evaluation Results for Rank-1 Accuracy (\%,$\uparrow$) on X-Palm Dataset. }
    \label{tab:rank1_cdcs_evaluation}
    \resizebox{\textwidth}{!}{
    \begin{tabular}{l|cccccccccc|c}
        \toprule
        \textbf{Test Domain} & \textbf{CompNet} & \textbf{PPNet} & \textbf{CCNet} & \textbf{CO3Net} & \textbf{SF2Net} & \textbf{PalmBridge} & \textbf{TSCAN} & \textbf{GIFT} & \textbf{ConvNeXt} & \textbf{DINOv2} & \textbf{Average} \\
        \midrule
        Scanner & \underline{69.84} & 59.72 & 68.55 & 62.81 & 67.78 & \textbf{70.35} & 26.21 & 38.22 & 19.21 & 34.39 & 51.71 \\
        Smartphone & \textbf{66.78} & 42.61 & 58.64 & 44.89 & \underline{60.30} & 56.27 & 35.99 & 35.29 & 30.96 & 36.22 & 46.79 \\
        Wet \& Text & 76.24 & 63.47 & \textbf{83.69} & 67.20 & 72.52 & \underline{79.43} & 40.07 & 58.69 & 52.31 & 78.37 & 67.20 \\
        Wet \& RND & 63.37 & 51.65 & 67.76 & 52.33 & 66.63 & \textbf{74.74} & 31.68 & 65.24 & 62.39 & \underline{69.03} & 60.48 \\
        RND \& Text & 78.24 & 65.61 & \underline{82.45} & 64.03 & 73.50 & \textbf{89.82} & 47.02 & 52.81 & 60.00 & 82.11 & 69.56 \\
        SF \& Roll & 45.25 & 28.49 & 49.72 & 37.99 & 51.39 & 74.86 & 51.58 & 63.87 & \underline{77.10} & \textbf{84.36} & 56.46 \\
        JF \& Pitch & 63.47 & 42.73 & 69.32 & 51.24 & 68.79 & \textbf{86.35} & 56.74 & 76.95 & \underline{82.27} & 82.27 & 68.01 \\
        BF \& Far & 89.65 & 80.56 & \underline{93.40} & 74.15 & 89.12 & \textbf{97.68} & 66.13 & 92.34 & 85.38 & 88.06 & 85.65 \\
        Roll \& Close & 55.31 & 37.99 & 55.86 & 41.90 & 55.31 & \underline{77.09} & 55.86 & 64.80 & 73.37 & \textbf{81.19} & 59.87 \\
        Far \& JF & 85.26 & 82.10 & 89.99 & 77.89 & 92.10 & \textbf{97.89} & 67.19 & \underline{92.98} & 85.62 & 89.30 & 86.03 \\
        FO \& SF & 96.49 & 84.91 & \underline{97.01} & 92.28 & 95.96 & \textbf{99.12} & 58.95 & 94.21 & 95.08 & 94.56 & 90.86 \\
        Roll \& Pitch & 39.93 & 31.45 & 42.19 & 30.89 & 41.06 & \underline{71.00} & 45.95 & 54.99 & 67.23 & \textbf{75.71} & 50.04 \\
        \midrule
        \textbf{Average} & 69.15 & 55.94 & 71.55 & 58.13 & 69.54 & \textbf{81.22} & 48.61 & 65.87 & 65.91 & \underline{74.63} & -- \\
        \bottomrule
    \end{tabular}
    }
\end{table*}

\begin{table*}[h]
    \centering
    \caption{Closed-Set Cross-Domain Evaluation Results for EER (\%,$\downarrow$) on X-Palm Dataset. }
    \label{tab:eer_cdcs_evaluation}
    \resizebox{\textwidth}{!}{
    \begin{tabular}{l|cccccccccc|c}
        \toprule
        \textbf{Test Domain} & \textbf{CompNet} & \textbf{PPNet} & \textbf{CCNet} & \textbf{CO3Net} & \textbf{SF2Net} & \textbf{PalmBridge} & \textbf{TSCAN} & \textbf{GIFT} & \textbf{ConvNeXt} & \textbf{DINOv2} & \textbf{Average} \\
        \midrule
        Scanner & \underline{28.33} & 30.74 & 28.35 & 28.97 & 29.19 & \textbf{25.37} & 47.95 & 40.21 & 36.32 & 33.04 & 32.85 \\
        Smartphone & \textbf{20.70} & 29.84 & 22.03 & 27.55 & 22.70 & \underline{21.64} & 29.91 & 31.13 & 27.09 & 23.31 & 25.59 \\
        Wet \& Text & 6.33 & 11.37 & \textbf{4.35} & 8.71 & 8.87 & 7.64 & 17.86 & 15.96 & 13.98 & \underline{5.92} & 10.10 \\
        Wet \& RND & 11.63 & 17.90 & 10.17 & 14.22 & 11.91 & \underline{8.59} & 19.41 & 12.20 & 9.60 & \textbf{8.13} & 12.38 \\
        RND \& Text & 13.13 & 18.68 & 9.97 & 17.03 & 15.04 & \underline{8.54} & 18.49 & 21.51 & 10.70 & \textbf{6.45} & 13.95 \\
        SF \& Roll & 17.19 & 21.77 & 13.28 & 18.86 & 15.17 & 6.57 & 13.41 & 14.90 & \underline{4.70} & \textbf{3.54} & 12.94 \\
        JF \& Pitch & 10.52 & 16.21 & 10.68 & 12.64 & 8.87 & \underline{5.20} & 11.78 & 6.95 & 6.21 & \textbf{4.61} & 9.37 \\
        BF \& Far & \underline{2.31} & 4.93 & 2.96 & 5.52 & 3.60 & \textbf{0.71} & 9.48 & 2.52 & 3.17 & 2.63 & 3.78 \\
        Roll \& Close & 14.20 & 21.19 & 15.21 & 17.58 & 18.22 & 7.50 & 15.56 & 15.64 & \underline{7.45} & \textbf{4.10} & 13.66 \\
        Far \& JF & 3.43 & 5.33 & 2.06 & 5.77 & 2.01 & \textbf{1.22} & 8.56 & \underline{1.73} & 3.34 & 2.81 & 3.63 \\
        FO \& SF & 1.23 & 4.39 & 1.54 & 2.28 & 1.02 & \underline{0.70} & 8.42 & 2.13 & \textbf{0.62} & 1.67 & 2.40 \\
        Roll \& Pitch & 19.30 & 22.60 & 14.97 & 20.06 & 17.69 & 11.02 & 16.80 & 17.58 & \underline{9.80} & \textbf{5.84} & 15.57 \\
        \midrule
        \textbf{Average} & 12.36 & 17.08 & 11.30 & 14.93 & 12.86 & \underline{8.72} & 18.14 & 15.20 & 11.08 & \textbf{8.50} & -- \\
        \bottomrule
    \end{tabular}
    }
\end{table*}


\textbf{Open-Set Cross-Domain Evaluation}
The initial two rows of Tables~\ref{tab:rank1_openset_cross_domain} 
and~\ref{tab:eer_openset_cross_domain} correspond to the cross-setting authentication scenarios, where models are trained on one acquisition setting and evaluated on the other, following the same protocol as the closed-set experiment but with disjoint train and test identities. The remaining rows confirm that \textit{Roll}, \textit{Pitch}, \textit{Wet}, \textit{Text}, and \textit{Random} domains consistently present the greatest challenge across all baselines, consistent with the findings of the closed-set cross-domain experiment. The performance degradation is further amplified in 
the open-set scenario, as models must simultaneously handle the challenging visual conditions described above and generalize to entirely unseen identities, compounding the difficulty of authentication.


\begin{table*}[h]
    \centering
    \caption{Open-Set Cross-Domain Evaluation Results for Rank-1 Accuracy (\%,$\uparrow$) on X-Palm Dataset. }
    \label{tab:rank1_openset_cross_domain}
    \resizebox{\textwidth}{!}{
    \begin{tabular}{l|cccccccccc|c}
        \toprule
        \textbf{Test Domain} & \textbf{CompNet} & \textbf{PPNet} & \textbf{CCNet} & \textbf{CO3Net} & \textbf{SF2Net} & \textbf{PalmBridge} & \textbf{ConvNeXt} & \textbf{DINOv2} & \textbf{ArcFace} & \textbf{MagFace} & \textbf{Average} \\
        \midrule
        Scanner & \underline{76.45} & 59.78 & 73.78 & 74.78 & \textbf{79.78} & 73.78 & 35.77 & 43.44 & 40.78 & 45.12 & 60.35 \\
        Smartphone & \textbf{68.00} & 38.25 & 57.27 & 47.45 & 61.25 & \underline{64.93} & 44.68 & 53.88 & 62.37 & 54.40 & 55.25 \\
        Wet \& Text & 50.88 & 35.09 & 45.61 & 37.72 & 42.98 & \underline{64.03} & 26.32 & 50.00 & \textbf{68.42} & 55.26 & 47.63 \\
        Wet \& RND & \underline{43.86} & 30.70 & \underline{43.86} & \underline{43.86} & 30.70 & 41.23 & 21.93 & 40.35 & \textbf{48.25} & 37.72 & 38.25 \\
        RND \& Text & 23.68 & 26.31 & 36.83 & 18.41 & 15.78 & 39.47 & 18.42 & 36.84 & \textbf{52.63} & \underline{44.73} & 31.31 \\
        SF \& Roll & 24.56 & 16.67 & 29.83 & 21.93 & 24.56 & 45.62 & \underline{57.02} & \textbf{59.65} & 56.14 & 40.35 & 37.63 \\
        JF \& Pitch & \textbf{43.85} & 22.80 & 25.43 & 28.06 & 28.06 & 30.70 & 29.83 & \underline{35.09} & 32.46 & \underline{35.09} & 31.14 \\
        BF \& Far & \textbf{78.07} & 49.13 & 67.55 & 49.13 & 70.18 & \underline{72.81} & 66.67 & 69.30 & 68.42 & 52.63 & 64.39 \\
        Roll \& Close & 51.75 & 33.33 & 35.96 & 38.59 & 54.38 & 51.75 & 46.50 & \underline{54.39} & \textbf{56.14} & 45.62 & 46.84 \\
        Far \& JF & 55.27 & 34.21 & 63.16 & 47.37 & 52.63 & 63.16 & \textbf{71.93} & \textbf{71.93} & \underline{64.03} & 53.50 & 57.72 \\
        FO \& SF & 76.32 & 34.22 & 63.16 & 55.27 & 60.53 & \textbf{78.95} & 58.77 & 71.93 & \underline{77.19} & 53.50 & 62.98 \\
        Roll \& Pitch & 28.95 & 21.05 & 36.84 & 23.68 & 31.58 & 44.74 & \underline{47.37} & \textbf{57.89} & 44.74 & \underline{47.37} & 38.42 \\
        \midrule
        \textbf{Average} & 51.80 & 33.46 & 48.27 & 40.52 & 46.03 & \underline{55.93} & 43.77 & 53.72 & \textbf{55.96} & 47.11 & -- \\
        \bottomrule
    \end{tabular}
    }
\end{table*}

\begin{table*}[h]
    \centering
    \caption{Open-Set Cross-Domain Evaluation Results for EER (\%,$\downarrow$) on X-Palm Dataset.}
    \label{tab:eer_openset_cross_domain}
    \resizebox{\textwidth}{!}{
    \begin{tabular}{l|cccccccccc|c}
        \toprule
        \textbf{Test Domain} & \textbf{CompNet} & \textbf{PPNet} & \textbf{CCNet} & \textbf{CO3Net} & \textbf{SF2Net} & \textbf{PalmBridge} & \textbf{ConvNeXt} & \textbf{DINOv2} & \textbf{ArcFace} & \textbf{MagFace} & \textbf{Average} \\
        \midrule
        Scanner & 28.02 & 30.28 & \underline{27.94} & 28.30 & 29.91 & \textbf{24.45} & 37.35 & 34.68 & 38.00 & 37.49 & 31.64 \\
        Smartphone & \textbf{22.98} & 35.18 & 25.87 & 29.33 & 26.12 & \underline{23.88} & 28.00 & 25.58 & 26.84 & 29.73 & 27.35 \\
        Wet \& Text & \textbf{14.04} & 18.51 & 18.25 & 21.93 & 19.56 & \underline{14.03} & 25.44 & 14.91 & 15.91 & 20.81 & 18.34 \\
        Wet \& RND & 22.73 & 37.47 & 20.63 & 20.10 & 27.73 & \underline{17.54} & 25.50 & 19.32 & \textbf{15.79} & 21.05 & 22.79 \\
        RND \& Text & 25.62 & 33.52 & 26.68 & 30.67 & 29.05 & \textbf{17.73} & 28.73 & 21.41 & \underline{20.17} & 24.87 & 25.85 \\
        SF \& Roll & 23.77 & 31.93 & 23.25 & 23.56 & 23.77 & 14.60 & \underline{14.03} & \textbf{11.40} & 15.98 & 25.94 & 20.82 \\
        JF \& Pitch & 26.17 & 30.91 & 26.96 & 36.70 & 24.60 & \textbf{18.28} & 24.11 & \underline{21.48} & 25.61 & 23.55 & 25.84 \\
        BF \& Far & 11.34 & 21.34 & 12.65 & 12.26 & 11.34 & \underline{8.70} & 12.78 & 11.85 & 13.39 & \textbf{8.42} & 12.41 \\
        Roll \& Close & 25.22 & 28.45 & 28.72 & 25.29 & 23.19 & 14.91 & 21.02 & \textbf{11.85} & \underline{14.04} & 24.57 & 21.73 \\
        Far \& JF & 14.35 & 22.95 & 16.64 & 19.32 & 18.21 & 9.79 & \underline{8.22} & \textbf{7.94} & 12.26 & 20.15 & 14.98 \\
        FO \& SF & \textbf{6.14} & 18.25 & 11.41 & 13.26 & 8.77 & \underline{6.12} & 11.87 & 9.74 & 10.69 & 17.59 & 11.39 \\
        Roll \& Pitch & 20.79 & 28.23 & 18.76 & 20.86 & 20.07 & \underline{15.60} & 16.95 & \textbf{13.54} & 20.18 & 26.09 & 20.11 \\
        \midrule
        \textbf{Average} & 20.10 & 28.09 & 21.48 & 23.46 & 21.86 & \textbf{15.47} & 21.17 & \underline{16.97} & 19.07 & 23.36 & -- \\
        \bottomrule
    \end{tabular}
    }
\end{table*}

\section{Limitations, Ethical Considerations and Societal Impact}

\textbf{Data Limitations}
\label{limit}
The primary limitation of the current version of X-Palm is its restricted scale, which makes it challenging to fully capture the breadth of real-world compound variations. Training more robust models capable of simultaneously handling diverse challenges requires a larger dataset with 
more identities, more samples per identity, and crucially, more iterations per variation type to better represent the full distribution of each challenge. As future work, we plan to extend X-Palm by enrolling additional participants across broader demographic groups and collecting denser per-identity samples for each acquisition condition, providing a more comprehensive benchmark for compound in-the-wild palmprint authentication.

\textbf{Ethical Considerations}
\label{ethics}
To safeguard participant privacy, all personally identifiable information was removed during pre-processing and replaced with anonymized subject identifiers, and the raw data is stored on access-controlled, encrypted servers maintained by the host institution. Dataset access is restricted to bona fide researchers for non-commercial academic use: prospective users must digitally sign an End User License Agreement (EULA) prohibiting redistribution, re-identification attempts, and commercial exploitation, after which approved requests are granted a time-limited secure download link within 24 hours. The complete access policy, EULA, and a datasheet documenting collection and intended use are published in our repository for review~\footnote{Please submit your request via this link to access the dataset:\url{https://forms.office.com/r/QvwACjaC0c}}.



\textbf{Societal Impact}
\label{impact}
X-Palm advances research in palmprint biometrics with applications in mobile banking, enterprise access control, healthcare systems, and personal device login. By addressing multispectral-to-smartphone Authentication and incorporating demographic, hardware, and condition diversities in X-Palm, it supports the development of fairer and more robust recognition systems. However, it carries risks of misuse for unauthorized surveillance or mass identification. To mitigate these risks, X-Palm is released under a strict non-commercial EULA, restricting usage to academic research, and no personally identifiable information is distributed with the dataset.

\section{Conclusion}
We introduced X-Palm, a cross-domain palmprint dataset of 6,006 images from 103 individuals. By pairing controlled multispectral scanner images with unconstrained smartphone images from the same identities, X-Palm provides the first benchmark explicitly targeting the discrepancy between enrollment and inference settings. Additionally, the smartphone data encompasses a broad spectrum of in-the-wild variability. Comprehensive experiments across 12 baselines and three evaluation protocols demonstrate that X-Palm is a significantly more challenging benchmark than existing datasets, while also serving as a superior training source that enhances cross-domain generalization. X-Palm can be used in developing more robust and generalizable palmprint authentication systems, and open new research opportunities in domain adaptation under compound variability and privacy-preserving mobile biometrics.

\bibliographystyle{ieeetr}
\bibliography{refs} 


\newpage
\appendix

\section{Appendix / supplemental material}

\subsection{Related Works}

\textbf{In-the-Wild Palmprint Authentication} CASIA-MS \cite{casiams} introduced multi-spectral imaging to capture hand images in varying spectra and enhance anti-spoofing capabilities. It is a contactless
multi-spectral palmprint dataset comprising
72, 000 images captured from 200 hands (100 subjects)
across six different spectra in a controlled setting. 
Tongji Mobile Palmprint Dataset (MPD-v2) \cite{mpd} is a large-scale contactless palmprint dataset comprising 16, 000 images captured from 400 hands
(200 subjects), collected using two smartphones (Huawei and Xiaomi) with variations in the background and lighting conditions.
Similarly, XJTU-UP \cite{xjtu}, contained 20,000 samples from 200 hands, specifically targets variations in acquisition device and lighting conditions. Palm images are collected by five different smartphones (Samsung Galaxy Note5, LG G4, MI 8, HUAWEI Mate 8, iPhone 6S) in two different lighting conditions (natural and flash light), simulating the heterogeneity of hardware and illumination encountered in actual deployments. However, in a real-life scenario, users capture their palm images with a wider spectrum of devices and under more challenging conditions.  
In NTU-PI \cite{ntu}, about 7,781 palm images (2035 hands) are collected from the Internet with different yet limited variations (about 4 by average) for each hand. 

\textbf{Cross-Domain and Open-Set Palmprint Recognition.}
Several works have addressed cross-domain generalization and open-set verification in the palmprint literature. For \textit{domain adaptation}, TSCAN~\cite{tscan} employs a teacher-student co-learning framework to adapt palmprint models across devices, DFMT~\cite{dfmt} leverages multi-teacher distillation for multi-target cross-dataset recognition, and LSFM~\cite{lsfm} proposes light style and feature matching for efficient cross-domain palmprint recognition. For \textit{domain generalization}, PDFG~\cite{pdfg} learns to generalize to unseen datasets, \cite{uaa} proposes unified adversarial augmentation to improve palmprint recognition robustness, PalmRSS~\cite{palmrss} addresses single-source domain generalization for palm biometrics, and GIFT~\cite{gift} generates stylized features for single-source cross-dataset recognition without access to the target domain. For \textit{open-set verification}, PalmBridge~\cite{palmbridge} introduces a plug-and-play feature alignment framework for open-set palmprint verification, and Palm-ID~\cite{palmid} employs multiscale multimodel embeddings for mobile contactless palmprint recognition. 

\textbf{Cross-Setting Authentication} BioSecure \cite{ortega2010biosecure} collected multimodal biometric data from more than 600 individuals in three distinct acquisition scenarios, controlled Desktop, unsupervised Internet, and unconstrained Mobile, enabling the evaluation of performance changes in different hardware settings. In the fingerprint domain, \cite{lin2018matching} targeted the severe modality shift between traditional hardware and modern touchless sensors by collecting paired contact-based and contactless samples from 300 individuals. RidgeBase \cite{jawade2022ridgebase} pushed the boundaries of contact-to-contactless matching by collecting over 15,000 paired fingerprint images from 88 individuals, specifically capturing the massive domain gap between high-quality flatbed scanner enrollment and in-the-wild smartphone verification under diverse lighting and background conditions. Despite these contributions in other modalities, the field of palmprint authentication still lacks a dataset addressing this cross-setting challenge. 


\subsection{Implementation Details.}
\label{details}
All experiments were conducted on an NVIDIA RTX A6000 (48\,GB). Experiments are repeated for 3 runs with independent seeds. All baselines were trained for 200 epochs with a batch size of 64 and input images resized to $112 \times 112$. Palmprint-specific models (CCNet, CO3Net, CompNet, PPNet, SF2Net, and PalmBridge) were optimized with Adam ($\text{lr} = 10^{-3}$, step LR decay). Pretrained vision models (ConvNeXtV2-Tiny and DINOv2 ViT-S/14) were fine-tuned with AdamW ($\text{lr} = 10^{-3}$, cosine annealing), keeping all but the last stage or last two transformer blocks frozen. ArcFace iResNet100 (pretrained on Glint360K) and MagFace iResNet100 (pretrained on MS1MV2) were fine-tuned with AdamW ($\text{lr} = 10^{-4}$, weight decay $= 5 \times 10^{-4}$, cosine annealing), with the first 75\% of parameter tensors frozen; MagFace retained its original adaptive margin formulation ($m_l = 0.45$, $m_u = 0.80$, $s = 64$, $\lambda_g = 20$). TSCAN followed a two-phase protocol with EMA updates (decay $= 0.999$) and pseudo-label filtering (threshold $= 0.70$). All models were trained with the same augmentation policy, contrast jitter, random crop, perspective distortion, and edge-anchored rotation. Matching was performed via cosine similarity on L2-normalized embeddings. To ensure fair comparison, gallery/probe splits were generated once with a fixed seed and shared across all models, and initial weights were cached after the first run and reused identically across all train--test combinations.

\subsection{Additional Experimental Results}
\label{sec:appendix_results}
\textbf{Leave-One-Out Cross-Dataset.} In this experiment, models are trained on a combined set of $K-1$ datasets and evaluated on the remaining held-out dataset to assess cross-dataset generalization. As shown in Table~\ref{tab:leave_one_out}, X-Palm consistently emerges as the most challenging benchmark, yielding the highest average EER (35.05\%) and lowest average Rank-1 accuracy (61.45\%). While models generalize well to controlled environments like XJTU-UP (EER=8.39\%) and CASIA-MS (EER=17.62\%), they fail to bridge the severe domain gap introduced by X-Palm's unconstrained compound variations. 
Among the baselines models, palmprint-specific architectures (PalmBridge, CompNet) and face-pretrained models (ArcFace, MagFace) demonstrate superior robustness compared to general vision models (ConvNeXt, DINOv2). Nevertheless, all evaluated architectures suffer a severe performance collapse on X-Palm, which proves that simply scaling training data cannot overcome cross-setting and cross domain challenges, motivating the need for novel methods explicitly designed for in-the-wild compound variability.

\begin{table*}[h]
  \centering
  \caption{Leave-One-Out Evaluation Results}
  \label{tab:leave_one_out}
  \resizebox{\textwidth}{!}{%
  \begin{tabular}{l cccccccc | cc}
    \toprule
    \multirow{2}{*}{\textbf{Model \textbackslash{} Held-Out Data}} & \multicolumn{2}{c}{\textbf{X-Palm}} & \multicolumn{2}{c}{\textbf{CASIA-MS}} & \multicolumn{2}{c}{\textbf{MPD-v2}} & \multicolumn{2}{c|}{\textbf{XJTU}} & \multicolumn{2}{c}{\textbf{Average}} \\
    \cmidrule(lr){2-3} \cmidrule(lr){4-5} \cmidrule(lr){6-7} \cmidrule(lr){8-9} \cmidrule(l){10-11}
    & \textbf{EER} & \textbf{Rank-1} & \textbf{EER} & \textbf{Rank-1} & \textbf{EER} & \textbf{Rank-1} & \textbf{EER} & \textbf{Rank-1} & \textbf{EER} & \textbf{Rank-1} \\
    \midrule
    CompNet    & 35.02 & 74.46 & 10.19 & 99.61 & 19.80 & 89.16 & 3.30  & 99.72 & \underline{17.08} & \textbf{90.74} \\
    PPNet      & 36.83 & 55.56 & 20.97 & 94.32 & 28.69 & 80.83 & 11.41 & 98.63 & 24.48 & 82.34 \\
    CCNet      & 34.74 & 66.60 & 12.73 & 98.68 & 20.42 & 87.65 & 4.78  & 99.60 & 18.17 & 88.13 \\
    CO3Net     & 35.44 & 65.69 & 13.52 & 98.60 & 21.71 & 86.28 & 5.28  & 99.36 & 18.99 & 87.51 \\
    SF2Net     & 36.49 & 71.60 & 10.82 & 99.57 & 19.75 & 87.87 & 3.13  & 99.80 & 17.55 & 89.71 \\
    PalmBridge & 31.55 & 73.71 & 14.32 & 97.24 & 16.65 & 91.52 & 4.93  & 99.44 & \textbf{16.86} & \underline{90.48} \\
    ConvNeXt   & 33.92 & 31.61 & 26.81 & 69.30 & 23.21 & 69.69 & 14.99 & 87.26 & 24.73 & 64.47 \\
    DINOv2     & 33.44 & 43.88 & 31.17 & 56.85 & 22.30 & 75.96 & 16.90 & 86.25 & 25.95 & 65.74 \\
    ArcFace    & 37.83 & 67.12 & 18.49 & 98.64 & 21.65 & 89.34 & 11.49 & 98.87 & 22.37 & 88.49 \\
    MagFace    & 35.22 & 64.26 & 17.14 & 98.75 & 22.75 & 89.42 & 7.66  & 99.44 & 20.69 & 87.97 \\
    \midrule
    \textbf{Average} & 35.05 & 61.45 & \underline{17.62} & \underline{91.16} & 21.69 & 84.77 & \textbf{8.39} & \textbf{96.84} & -- & -- \\
    \bottomrule
  \end{tabular}%
  }
\end{table*}

\subsection{Experiment Statistical Significance}
\label{stat_sig}

Due to space constraints, detailed results are reported for three representative baselines per protocol, PalmBridge, ArcFace, and DINOv2 for the cross-dataset and open-set protocols, and PalmBridge, GIFT, and DINOv2 for the closed-set protocol, though all 12 baselines are run three times with independent seeds. Tables~\ref{tab:significance_cross_dataset},
\ref{tab:significance_closed_set}, and~\ref{tab:significance_open_set} report mean\,$\pm$\,std over these three runs. The $\pm$ values denote one standard deviation (1-sigma), not standard error of the mean. No assumption of normality is made given the small number of repetitions; intervals are therefore reported as descriptive spread rather than formal confidence intervals.
Error bars capture variability across runs that differ in: (1)~random data sampling, which images are selected per dataset split; (2)~gallery/probe assignment; and (3)~model weight initialisation and training stochasticity (batch ordering, dropout).

\textbf{Cross-Dataset.}
As shown in Table~\ref{tab:significance_cross_dataset}, models show strong aggregate reproducibility across seeds, with mean cell EER standard deviations of 0.52 (PalmBridge), 0.63 (ArcFace), and 0.96 (DINOv2). PalmBridge is most stable when XJTU is the training source (avg EER std $= 0.32$) and most volatile at MPD-v2$\to$CASIA-MS (1.53) and CASIA-MS$\to$X-Palm (1.44), where source and target datasets differ most in acquisition complexity. ArcFace achieves lower mean R1 std than PalmBridge (1.15 vs.\ 1.36). DINOv2 is the most volatile overall, with XJTU as training source producing the largest R1 instability (avg std $= 4.79$).
\begin{table*}[t]
  \centering
  \caption{Cross-Dataset Evaluation: Mean\,$\pm$\,Std of EER\,(\%,\,$\downarrow$) and
           Rank-1\,(\%,\,$\uparrow$) over three independent runs}
  \label{tab:significance_cross_dataset}
  \resizebox{\textwidth}{!}{%
  \begin{tabular}{l l cc cc cc cc}
    \toprule
    \multirow{2}{*}{\textbf{Model}}
      & \multirow{2}{*}{\textbf{Train\,\textbackslash\,Test}}
      & \multicolumn{2}{c}{\textbf{X-Palm}}
      & \multicolumn{2}{c}{\textbf{CASIA-MS}}
      & \multicolumn{2}{c}{\textbf{MPD-v2}}
      & \multicolumn{2}{c}{\textbf{XJTU}} \\
    \cmidrule(lr){3-4}\cmidrule(lr){5-6}\cmidrule(lr){7-8}\cmidrule(lr){9-10}
    & & \textbf{EER} & \textbf{Rank-1}
      & \textbf{EER} & \textbf{Rank-1}
      & \textbf{EER} & \textbf{Rank-1}
      & \textbf{EER} & \textbf{Rank-1} \\
    \midrule
    \multirow{4}{*}{\textbf{PalmBridge}}
    & X-Palm   & $23.31\pm0.45$ & $79.49\pm2.30$ & $16.28\pm0.35$ & $89.75\pm2.49$ & $19.58\pm0.30$ & $89.15\pm1.09$ & $7.98\pm0.47$  & $98.25\pm0.32$ \\
    & CASIA-MS & $35.06\pm1.44$ & $63.00\pm3.79$ & $8.75\pm0.17$  & $99.28\pm0.24$ & $21.07\pm0.71$ & $87.52\pm1.61$ & $7.97\pm0.41$  & $98.90\pm0.33$ \\
    & MPD-v2   & $33.23\pm0.41$ & $59.22\pm2.15$ & $20.06\pm1.53$ & $82.31\pm3.14$ & $14.89\pm0.39$ & $91.75\pm0.51$ & $8.12\pm0.32$  & $97.02\pm0.15$ \\
    & XJTU     & $33.82\pm0.73$ & $58.18\pm0.58$ & $18.14\pm0.08$ & $90.03\pm2.29$ & $19.50\pm0.07$ & $86.44\pm0.40$ & $5.38\pm0.42$  & $99.65\pm0.44$ \\
    \midrule
    \multirow{4}{*}{\textbf{ArcFace}}
    & X-Palm   & $26.54\pm1.24$ & $82.52\pm0.13$ & $19.03\pm1.43$ & $96.59\pm0.35$ & $25.55\pm0.82$ & $85.10\pm0.99$ & $12.79\pm0.99$ & $98.28\pm0.41$ \\
    & CASIA-MS & $48.09\pm0.21$ & $26.15\pm0.81$ & $10.21\pm0.32$ & $99.81\pm0.20$ & $34.88\pm0.25$ & $72.53\pm0.47$ & $42.34\pm0.70$ & $62.40\pm2.64$ \\
    & MPD-v2   & $44.01\pm0.32$ & $35.19\pm1.79$ & $22.06\pm0.78$ & $95.96\pm0.51$ & $17.17\pm0.18$ & $91.35\pm2.68$ & $46.05\pm0.81$ & $65.66\pm4.87$ \\
    & XJTU     & $39.38\pm1.19$ & $60.07\pm1.12$ & $19.96\pm0.03$ & $97.11\pm0.65$ & $23.16\pm0.31$ & $86.01\pm0.72$ & $5.86\pm0.54$  & $99.86\pm0.12$ \\
    \midrule
    \multirow{4}{*}{\textbf{DINOv2}}
    & X-Palm   & $20.74\pm0.30$ & $76.08\pm1.15$ & $28.11\pm0.77$ & $55.41\pm2.33$ & $20.90\pm1.42$ & $70.52\pm2.69$ & $18.30\pm2.19$ & $82.24\pm3.94$ \\
    & CASIA-MS & $35.62\pm0.49$ & $35.11\pm2.25$ & $15.55\pm0.70$ & $88.17\pm2.10$ & $24.23\pm0.90$ & $69.57\pm2.67$ & $21.23\pm0.25$ & $79.79\pm1.25$ \\
    & MPD-v2   & $34.28\pm0.40$ & $33.58\pm2.08$ & $32.24\pm1.16$ & $55.98\pm1.71$ & $12.54\pm0.57$ & $92.22\pm1.55$ & $18.20\pm0.98$ & $83.83\pm1.09$ \\
    & XJTU     & $33.09\pm1.86$ & $36.38\pm4.25$ & $29.76\pm1.05$ & $55.41\pm5.97$ & $21.97\pm1.99$ & $68.67\pm7.61$ & $11.59\pm0.41$ & $93.17\pm1.34$ \\
    \bottomrule
  \end{tabular}}
\end{table*}

\textbf{Closed-Set Cross-Domain.}
As depicted in Table \ref{tab:significance_closed_set}, models are substantially more reproducible in the closed-set setting than in the open-set, since the only sources of randomness are the gallery/probe split and model initialisation rather than identity composition. PalmBridge (mean EER std $= 0.59$) exhibits a clear two-tier stability structure: near-ceiling domains such as BF\,\&\,Far, Far\,\&\,JF, FO\,\&\,SF, and Scanner are nearly deterministic (EER std $\leq 0.31$), while geometrically demanding domains -- Roll\,\&\,Close (1.34), Roll\,\&\,Pitch (1.22), and SF\,\&\,Roll (1.11) carry the highest variance, as pose distortions make the visible palm region sensitive to which images fall into the gallery versus probe split. The Smartphone domain additionally shows elevated R1 std (5.20), driven by the cross-setting acquisition gap rather than pose alone. GIFT (mean EER std $= 1.24$) is the least stable of the three methods, with its Scanner variance (EER std $= 4.50$) far exceeding any other domain-model pair in the closed-set. DINOv2 (mean EER std $= 0.80$) sits between the two: stable on most domains but showing elevated variance on Wet\,\&\,Text (1.57) and SF\,\&\,Roll (1.29).

\begin{table*}[h]
  \centering
  \caption{Closed-Set Cross-Domain Evaluation: Mean\,$\pm$\,Std of EER\,(\%,\,$\downarrow$)
           and Rank-1\,(\%,\,$\uparrow$) over three independent runs.}
  \label{tab:significance_closed_set}
  \resizebox{\textwidth}{!}{%
  \begin{tabular}{l cc cc cc}
    \toprule
    \multirow{2}{*}{\textbf{Test Domain}}
      & \multicolumn{2}{c}{\textbf{PalmBridge}}
      & \multicolumn{2}{c}{\textbf{GIFT}}
      & \multicolumn{2}{c}{\textbf{DINOv2}} \\
    \cmidrule(lr){2-3}\cmidrule(lr){4-5}\cmidrule(lr){6-7}
    & \textbf{EER} & \textbf{Rank-1}
    & \textbf{EER} & \textbf{Rank-1}
    & \textbf{EER} & \textbf{Rank-1} \\
    \midrule
    Scanner       & $25.37\pm0.09$ & $70.35\pm1.00$ & $40.21\pm4.50$ & $38.22\pm6.77$ & $33.04\pm1.10$ & $34.39\pm6.32$ \\
    Smartphone    & $21.64\pm0.82$ & $56.27\pm5.20$ & $31.13\pm1.97$ & $35.29\pm5.42$ & $23.31\pm0.12$ & $36.22\pm3.30$ \\
    Wet \& Text   & $7.64\pm0.84$  & $79.43\pm1.34$ & $15.96\pm0.71$ & $58.69\pm0.31$ & $5.92\pm1.57$  & $78.37\pm1.87$ \\
    Wet \& RND    & $8.59\pm0.29$  & $74.74\pm2.41$ & $12.20\pm0.73$ & $65.24\pm0.40$ & $8.13\pm0.75$  & $69.03\pm1.17$ \\
    RND \& Text   & $8.54\pm0.13$  & $89.82\pm0.61$ & $21.51\pm0.17$ & $52.81\pm4.89$ & $6.45\pm0.32$  & $82.11\pm1.39$ \\
    SF \& Roll    & $6.57\pm1.11$  & $74.86\pm2.95$ & $14.90\pm0.32$ & $63.87\pm2.11$ & $3.54\pm1.29$  & $84.36\pm1.48$ \\
    JF \& Pitch   & $5.20\pm0.40$  & $86.35\pm1.34$ & $6.95\pm0.89$  & $76.95\pm2.02$ & $4.61\pm1.23$  & $82.27\pm1.34$ \\
    BF \& Far     & $0.71\pm0.31$  & $97.68\pm0.31$ & $2.52\pm0.25$  & $92.34\pm1.24$ & $2.63\pm1.01$  & $88.06\pm5.25$ \\
    Roll \& Close & $7.50\pm1.34$  & $77.09\pm2.96$ & $15.64\pm2.96$ & $64.80\pm2.90$ & $4.10\pm0.32$  & $81.19\pm2.76$ \\
    Far \& JF     & $1.22\pm0.18$  & $97.89\pm0.53$ & $1.73\pm0.26$  & $92.98\pm1.21$ & $2.81\pm0.31$  & $89.30\pm1.61$ \\
    FO \& SF      & $0.70\pm0.30$  & $99.12\pm0.30$ & $2.13\pm0.14$  & $94.21\pm1.39$ & $1.67\pm1.29$  & $94.56\pm2.12$ \\
    Roll \& Pitch & $11.02\pm1.22$ & $71.00\pm1.30$ & $17.58\pm2.01$ & $54.99\pm2.84$ & $5.84\pm0.32$  & $75.71\pm0.98$ \\
    \bottomrule
  \end{tabular}}
\end{table*}

\textbf{Open-Set Cross-Domain.}
As illustrated in Table \ref{tab:significance_closed_set}, the open-set setting introduces a different source of variance absent in the closed-set: which 20\% of identities are assigned to the disjoint test split. With only 41 test identities per domain condition, a single identity reassignment can shift performance substantially, and this is reflected directly in the numbers. PalmBridge's average EER std nearly quadruples relative to closed-set (2.15 vs.\ 0.59), and average R1 std nearly quadruples as well (6.31 vs.\ 1.69). The highest-variance cells are Far\,\&\,JF (EER std $= 4.07$), BF\,\&\,Far (3.00), and Roll\,\&\,Close (2.91) in EER, and Roll\,\&\,Close (R1 std $= 15.86$), Wet\,\&\,Text (8.46), and FO\,\&\,SF (7.90) in Rank-1 accuracy, the same domains that drive instability in the closed-set, but with substantially larger magnitudes. ArcFace (mean EER std $= 1.83$) and DINOv2 (mean EER std $= 1.79$) are both comparable to each other and slightly more stable than PalmBridge in EER. 

\begin{table*}[h]
  \centering
  \caption{Open-Set Cross-Domain Evaluation: Mean\,$\pm$\,Std of EER\,(\%,\,$\downarrow$)
           and Rank-1\,(\%,\,$\uparrow$) over three independent runs.}
  \label{tab:significance_open_set}
  \resizebox{\textwidth}{!}{%
  \begin{tabular}{l cc cc cc}
    \toprule
    \multirow{2}{*}{\textbf{Test Domain}}
      & \multicolumn{2}{c}{\textbf{PalmBridge}}
      & \multicolumn{2}{c}{\textbf{ArcFace}}
      & \multicolumn{2}{c}{\textbf{DINOv2}} \\
    \cmidrule(lr){2-3}\cmidrule(lr){4-5}\cmidrule(lr){6-7}
    & \textbf{EER} & \textbf{Rank-1}
    & \textbf{EER} & \textbf{Rank-1}
    & \textbf{EER} & \textbf{Rank-1} \\
    \midrule
    Scanner       & $24.45\pm0.64$  & $73.78\pm3.42$  & $38.00\pm0.61$ & $40.78\pm5.09$ & $34.68\pm0.64$ & $43.44\pm4.35$  \\
    Smartphone    & $23.88\pm1.17$  & $64.93\pm4.08$  & $26.84\pm0.85$ & $62.37\pm3.36$ & $25.58\pm1.05$ & $53.88\pm0.70$  \\
    Wet \& Text   & $14.04\pm1.52$  & $64.03\pm8.46$  & $15.91\pm2.64$ & $68.42\pm6.96$ & $14.91\pm1.52$ & $50.00\pm4.56$  \\
    Wet \& RND    & $17.54\pm2.50$  & $41.23\pm6.62$  & $15.79\pm2.63$ & $48.25\pm3.04$ & $19.32\pm0.86$ & $40.35\pm4.02$  \\
    RND \& Text   & $17.73\pm3.37$  & $39.47\pm2.64$  & $20.17\pm3.04$ & $52.63\pm7.90$ & $21.41\pm6.83$ & $36.84\pm16.00$ \\
    SF \& Roll    & $14.60\pm1.14$  & $45.62\pm4.02$  & $15.98\pm0.33$ & $56.14\pm5.48$ & $11.40\pm1.52$ & $59.65\pm8.04$  \\
    JF \& Pitch   & $18.28\pm2.42$  & $30.70\pm5.47$  & $25.61\pm4.77$ & $32.46\pm3.04$ & $21.48\pm2.02$ & $35.09\pm1.52$  \\
    BF \& Far     & $8.70\pm3.00$   & $72.81\pm5.48$  & $13.39\pm2.17$ & $68.42\pm6.96$ & $11.85\pm1.19$ & $69.30\pm4.02$  \\
    Roll \& Close & $14.91\pm2.91$  & $51.75\pm15.86$ & $14.04\pm1.52$ & $56.14\pm4.02$ & $11.85\pm1.31$ & $54.39\pm9.96$  \\
    Far \& JF     & $9.79\pm4.07$   & $63.16\pm9.12$  & $12.26\pm0.97$ & $64.03\pm7.60$ & $7.94\pm2.56$  & $71.93\pm8.04$  \\
    FO \& SF      & $6.14\pm1.52$   & $78.95\pm7.90$  & $10.69\pm1.89$ & $77.19\pm5.48$ & $9.74\pm1.35$  & $71.93\pm6.62$  \\
    Roll \& Pitch & $15.60\pm1.50$  & $44.74\pm2.63$  & $20.18\pm0.60$ & $44.74\pm6.96$ & $13.54\pm0.66$ & $57.89\pm2.64$  \\
    \bottomrule
  \end{tabular}}
\end{table*}

\subsection{Manual Annotation and ROI Extraction.}
\label{roi}

\begin{figure}[h]
    \centering
    \includegraphics[width=0.85\linewidth]{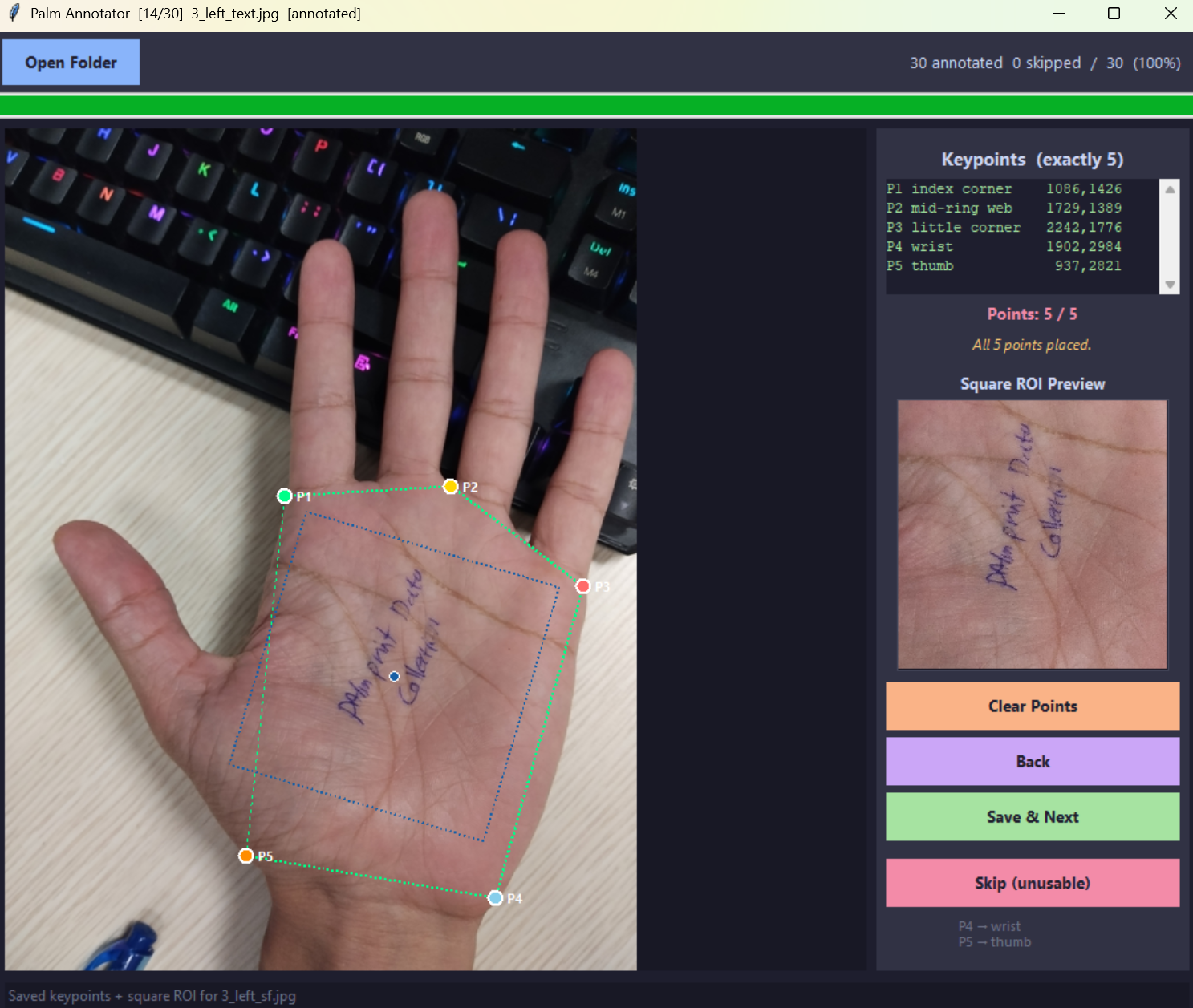}
    \caption{The custom palm annotation tool. An operator marks five 
    anatomical keypoints ($P_1$--$P_5$) on each image. The green 
    pentagon connects the keypoints, the blue dashed square shows the 
    extracted ROI boundary, and the right panel displays a live preview 
    of the extracted square ROI.}
    \label{fig:annotator}
\end{figure}

To ensure annotation quality and establish a ground truth for future development of automated ROI extraction methods, palm images are manually annotated using a custom-built annotation tool, as shown in Fig.~\ref{fig:annotator}. For each image, an operator marks five anatomical keypoints: $P_1$ (index finger base corner), $P_2$ (mid-ring finger web), $P_3$ (little finger base corner), $P_4$ (wrist), and $P_5$ (thumb base). These five points were selected to define a stable and repeatable coordinate frame that captures the principal discriminative region of the palm. Specifically, $P_1$ and $P_3$ define the upper boundary of the palm along the finger-base line, $P_4$ and $P_5$ anchor the lower wrist region, and $P_2$ provides a central reference along the transverse axis. Together, they unambiguously determine the palm extent regardless of hand orientation or distance from the camera.

The ROI is extracted following the method of ~\cite{roi}, adapted to our annotation scheme. Let $D = \|P_1 - P_3\|$ denote the Euclidean distance between the index and little finger corners, and $M = \frac{P_1 + P_3}{2}$ the midpoint of this axis. A unit vector $\hat{u} = \frac{P_3 - P_1}{D}$ is defined along the $P_1 \to P_3$ 
direction, and its perpendicular $\hat{p} = R_{90}\hat{u}$ is validated to point toward the wrist by checking its dot product with $P_4 - M$, flipping sign if necessary. The ROI centre is then placed at:
\begin{equation}
    O' = M + \alpha \cdot D \cdot \hat{p}
\end{equation}
where $\alpha = 0.45$ controls the offset from the finger-base line 
toward the palm centre. A square region of side $s = \beta \cdot D$ 
with $\beta = 0.85$ is extracted, aligned with $\hat{u}$ and $\hat{p}$, 
giving four corners:
\begin{align}
    C_\text{TL} &= O' - \tfrac{s}{2}\hat{u} - \tfrac{s}{2}\hat{p} \\
    C_\text{TR} &= O' + \tfrac{s}{2}\hat{u} - \tfrac{s}{2}\hat{p} \\
    C_\text{BR} &= O' + \tfrac{s}{2}\hat{u} + \tfrac{s}{2}\hat{p} \\
    C_\text{BL} &= O' - \tfrac{s}{2}\hat{u} + \tfrac{s}{2}\hat{p}
\end{align}
The ROI is then extracted via an affine transformation that maps this 
oriented square to a fixed $112 \times 112$ pixel output, preserving 
the spatial alignment across all captures regardless of hand orientation 
or scale.

\begin{figure*}[t] 
    \centering
    
    \begin{subfigure}{0.16\textwidth}
        \includegraphics[width=\linewidth]{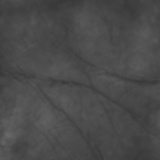}
        \caption{L\_001\_460}
    \end{subfigure}\hfill
    \begin{subfigure}{0.16\textwidth}
        \includegraphics[width=\linewidth]{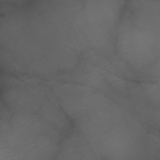}
        \caption{L\_001\_850}
    \end{subfigure}\hfill
    \begin{subfigure}{0.16\textwidth}
        \includegraphics[width=\linewidth]{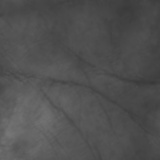}
        \caption{L\_001\_WHT}
    \end{subfigure}\hfill
    \begin{subfigure}{0.16\textwidth}
        \includegraphics[width=\linewidth]{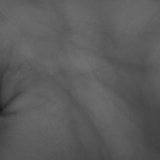}
        \caption{L\_002\_700}
    \end{subfigure}\hfill
    \begin{subfigure}{0.16\textwidth}
        \includegraphics[width=\linewidth]{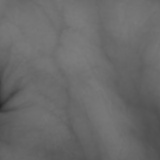}
        \caption{L\_002\_940}
    \end{subfigure}\hfill
    \begin{subfigure}{0.16\textwidth}
        \includegraphics[width=\linewidth]{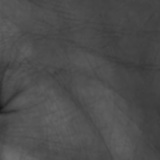}
        \caption{L\_002\_WHT}
    \end{subfigure}
    
    \vspace{-0.05cm} 
    
    \begin{subfigure}{0.16\textwidth}
        \includegraphics[width=\linewidth]{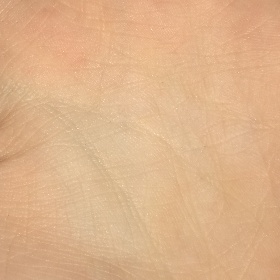}
        \caption{L\_001\_H\_F}
    \end{subfigure}\hfill
    \begin{subfigure}{0.16\textwidth}
        \includegraphics[width=\linewidth]{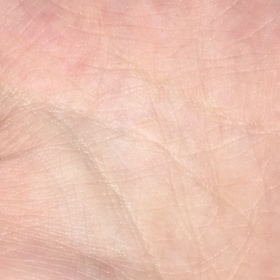}
        \caption{L\_001\_I\_F}
    \end{subfigure}\hfill
    \begin{subfigure}{0.16\textwidth}
        \includegraphics[width=\linewidth]{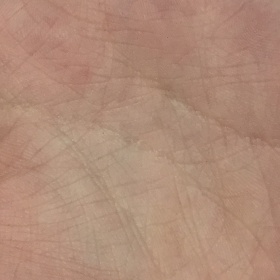}
        \caption{L\_001\_I\_N}
    \end{subfigure}\hfill
    \begin{subfigure}{0.16\textwidth}
        \includegraphics[width=\linewidth]{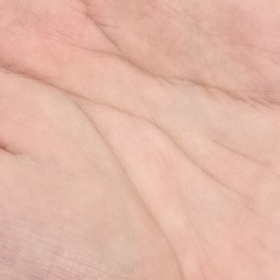}
        \caption{L\_002\_H\_N}
    \end{subfigure}\hfill
    \begin{subfigure}{0.16\textwidth}
        \includegraphics[width=\linewidth]{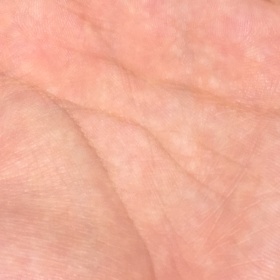}
        \caption{L\_002\_I\_F}
    \end{subfigure}\hfill
    \begin{subfigure}{0.16\textwidth}
        \includegraphics[width=\linewidth]{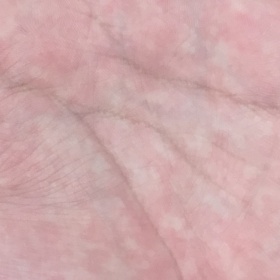}
        \caption{L\_002\_I\_N}
    \end{subfigure}
    
    \vspace{-0.05cm} 
    
    \begin{subfigure}{0.16\textwidth}
        \includegraphics[width=\linewidth]{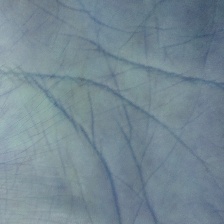}
        \caption{L\_001\_S1\_H}
    \end{subfigure}\hfill
    \begin{subfigure}{0.16\textwidth}
        \includegraphics[width=\linewidth]{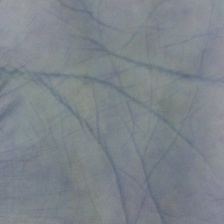}
        \caption{L\_001\_S1\_M}
    \end{subfigure}\hfill
    \begin{subfigure}{0.16\textwidth}
        \includegraphics[width=\linewidth]{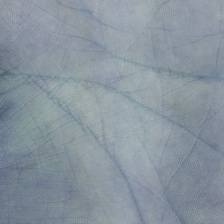}
        \caption{L\_001\_S1\_H}
    \end{subfigure}\hfill
    \begin{subfigure}{0.16\textwidth}
        \includegraphics[width=\linewidth]{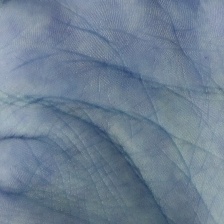}
        \caption{L\_003\_S1\_H}
    \end{subfigure}\hfill
    \begin{subfigure}{0.16\textwidth}
        \includegraphics[width=\linewidth]{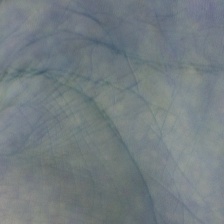}
        \caption{L\_003\_S1\_M}
    \end{subfigure}\hfill
    \begin{subfigure}{0.16\textwidth}
        \includegraphics[width=\linewidth]{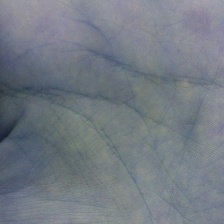}
        \caption{L\_003\_S2\_M}
    \end{subfigure}
    
    \vspace{-0.05cm} 
    
    \begin{subfigure}{0.16\textwidth}
        \includegraphics[width=\linewidth]{roi_samples/3_left_bf.jpg}
        \caption{L\_003\_bf}
    \end{subfigure}\hfill
    \begin{subfigure}{0.16\textwidth}
        \includegraphics[width=\linewidth]{roi_samples/3_left_text.jpg}
        \caption{L\_003\_text}
    \end{subfigure}\hfill
    \begin{subfigure}{0.16\textwidth}
        \includegraphics[width=\linewidth]{roi_samples/3_left_wet.jpg}
        \caption{L\_003\_wet}
    \end{subfigure}\hfill
    \begin{subfigure}{0.16\textwidth}
        \includegraphics[width=\linewidth]{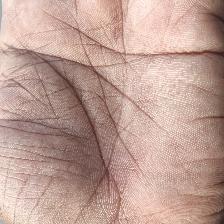}
        \caption{L\_005\_close}
    \end{subfigure}\hfill
    \begin{subfigure}{0.16\textwidth}
        \includegraphics[width=\linewidth]{roi_samples/5_left_far.jpg}
        \caption{L\_005\_far}
    \end{subfigure}\hfill
    \begin{subfigure}{0.16\textwidth}
        \includegraphics[width=\linewidth]{roi_samples/5_left_roll.jpg}
        \caption{L\_005\_roll}
    \end{subfigure}
    
    \caption{Visual comparison of palmprint samples across four different datasets: CASIA-MS (top row), XJTU (second row), MPD-v2 (third row), and X-Palm (bottom row).}
    \label{fig:dataset_comparison_grid}
\end{figure*}

Fig.\ref{fig:dataset_comparison_grid} provides a visual comparison of ROI samples extracted in four different datasets. To illustrate both intra-class and inter-class diversities, each row presents six samples, systematically partitioned into two distinct identities (the first three images belong to one identity, and the subsequent three to another). As observed in the first three rows, images collected under controlled conditions exhibit pronounced visual uniformity. Consequently, they demonstrate limited inter-class variance as different identities are captured under a similar homogeneous conditions, and with minimal intra-class domain shift where multiple captures from the same identity appear similar. In contrast, the bottom row highlights the highly unconstrained nature of the X-Palm dataset. By intentionally incorporating extensive variations such as varied hand poses, handwritten occlusions, surface moisture, and various camera distances and view points, X-Palm exhibits substantial intra-class domain shift. Furthermore, because participants captured their own data using diverse personal smartphones across uncontrolled environments, the dataset introduces significant inter-class variability, simulating the complex and compound challenges of real-world deployment.

\subsection{Dataset Demographics}
\label{sec:demographics}

\begin{figure}[h]
    \centering
    \includegraphics[width=\linewidth]{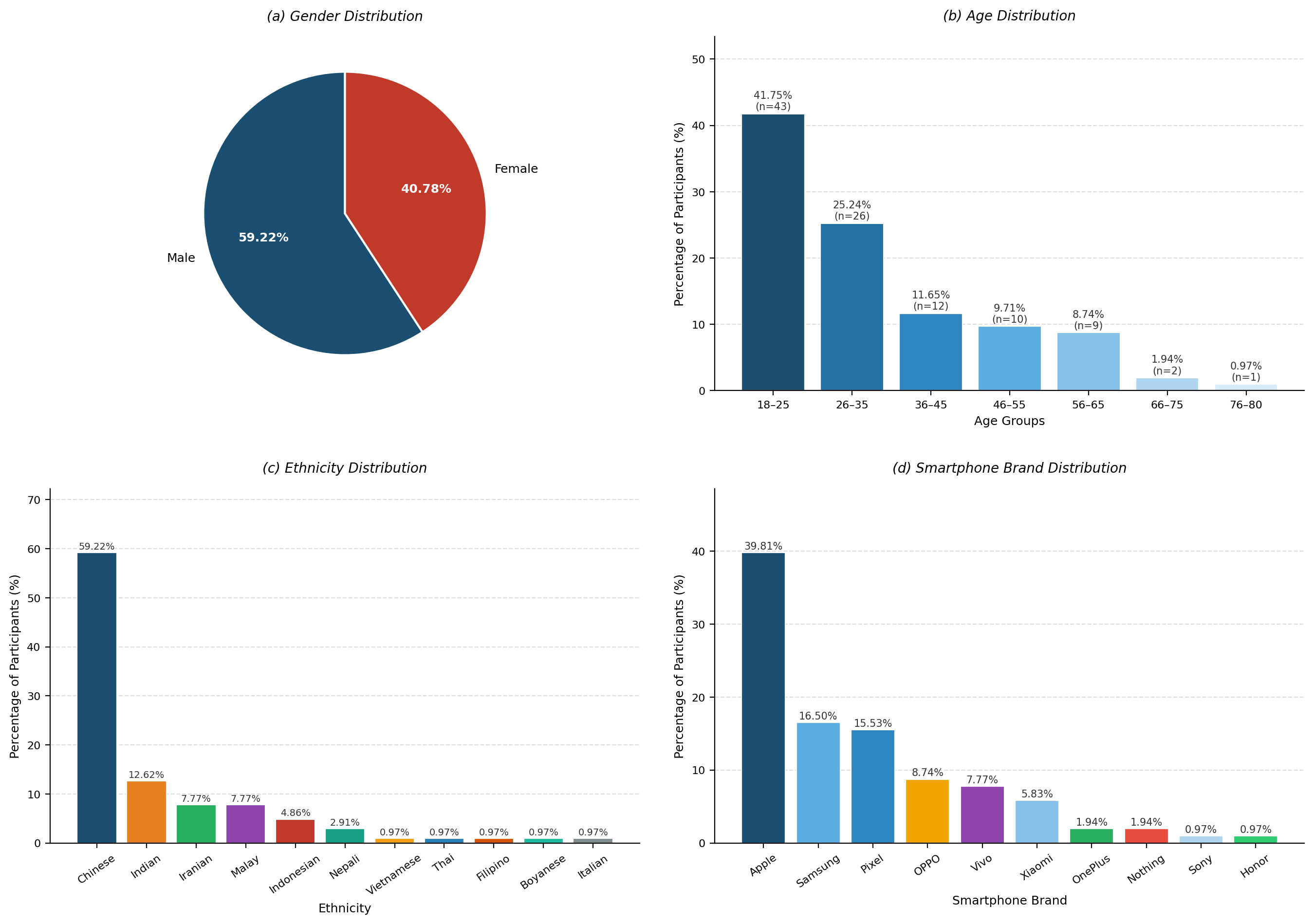}
    \caption{Demographic distribution of X-Palm participants: 
    (a) gender, (b) age groups, (c) ethnicity, and (d) smartphone brand.}
    \label{fig:demographics}
\end{figure}

\textbf{Gender and Age.}
The dataset comprises 103 participants: 61 male (59.22\%) and 42 female (40.78\%), aged 18--76 years. The age distribution is right-skewed, with the majority falling in the 18--25 group (41.75\%, $n=43$) and the 26--35 group (25.24\%, $n=26$), followed by 36--45 (11.65\%, $n=12$), 46--55 (9.71\%, $n=10$), 56--65 (8.74\%, $n=9$), 66--75 (1.94\%, $n=2$), and 76--80 (0.97\%, $n=1$), ensuring broad coverage across all adult age ranges.

\textbf{Ethnicity.}
Eleven distinct ethnic groups are represented. Chinese participants form the largest group (59.22\%, $n=61$), followed by Indian (12.62\%, $n=13$), Iranian (7.77\%, $n=8$), Malay (7.77\%, $n=8$), Indonesian (4.86\%, $n=5$), and Nepali (2.91\%, $n=3$). Vietnamese, Thai, Filipino, Boyanese, and Italian participants each contribute 0.97\% ($n=1$). This multi-ethnic composition reduces demographic bias and strengthens cross-population generalization.

\textbf{Smartphone Hardware.}
Captures span 10 device brands, led by Apple (39.81\%, $n=41$), Samsung (16.50\%, $n=17$), and Google Pixel (15.53\%, $n=16$), with the remainder distributed across OPPO (8.74\%, $n=9$), Vivo (7.77\%, $n=8$), Xiaomi (5.83\%, $n=6$), OnePlus (1.94\%, $n=2$), Nothing (1.94\%, $n=2$), Sony (0.97\%, $n=1$), and Honor (0.97\%, $n=1$). This hardware diversity introduces realistic variation in sensor characteristics, resolution, and image quality, making the dataset well-suited to benchmark palmprint recognition robustness under heterogeneous real-world acquisition conditions.

\end{document}